\documentclass[%
 aip,
 amsmath,amssymb,
 reprint,%
]{revtex4-1}
\usepackage{graphicx}
\usepackage{subfigure}
\usepackage{dcolumn}
\usepackage{bm}
\usepackage{color}
\usepackage[utf8]{inputenc}
\usepackage[T1]{fontenc}
\usepackage{mathptmx}
\usepackage{multirow}
\usepackage{float}
\usepackage{epstopdf}
\usepackage{titlesec}
\titleclass{\subsubsubsection}{straight}[\subsection]
\newcounter{subsubsubsection}[subsubsection]
\renewcommand\thesubsubsubsection{\thesubsubsection.\alph{subsubsubsection}}
\titleformat{\subsubsubsection}
 {\normalfont\normalsize\bfseries}{\thesubsubsubsection}{1em}{}
\titlespacing*{\subsubsubsection}
{0pt}{3.25ex plus 1ex minus .2ex}{1.5ex plus .2ex}

\begin{document}
\preprint{AIP/123-QED}

\title{Delineation of the flow and mixing induced by Rayleigh-Taylor instability through tracers}

\author{Ge Zhang}
\affiliation{Laboroatory of Computational Physics, Institute of Applied Physics and Computational Mathematics, Beijing 100088, China}
\author{Aiguo Xu}
\thanks{Corresponding author: xu\_aiguo@iapcm.ac.cn}
\affiliation{Laboroatory of Computational Physics, Institute of Applied Physics and Computational Mathematics, Beijing 100088, China}
\affiliation{State Key Laboratory of Explosion Science and Technology, Beijing Institute of Technology, Beijing 100081, China}
\affiliation{Center for Applied Physics and Technology, MOE  Key Center for High Energy Density Physics Simulations, College of Engineering, Peking University, Beijing 100871, China}

\author{Dejia Zhang}
\affiliation{State Key Laboratory for Geomechanics and Deep Underground Engineering, China University of Mining and Technology, Beijing 100083, China}

\author{Yingjun Li}
\affiliation{State Key Laboratory for Geomechanics and Deep Underground Engineering, China University of Mining and Technology, Beijing 100083, China}

\author{Huilin Lai}
\affiliation{Key Laboratory of Analytical Mathematics and Application in Fujian Province, College of Mathematics and Informatics, Fujian Normal University, Fuzhou 350007, China}

\author{Xiaomian Hu}
\affiliation{Laboroatory of Computational Physics, Institute of Applied Physics and Computational Mathematics, Beijing 100088, China}

\date{\today}

\begin{abstract}
Rayleigh-Taylor-instability(RTI) induced flow and mixing are of great importance in both nature and engineering scenarios. To capture the underpinning physics, tracers are introduced to make a supplement to  discrete Boltzmann simulation of RTI in compressible flows. Via marking two types of tracers with different colors, the tracer distribution provides a clear boundary of two fluids during the RTI evolution. Fine structures of the flow and thermodynamic nonequilibrium behavior around the interface in a miscible two-fluid system are delineated. Distribution of tracers in its velocity phase space makes a charming pattern showing quite dense information on the flow behavior, which opens a new perspective for analyzing and accessing significantly deep insights into the flow system. RTI mixing is further investigated via tracer defined local mixedness. The appearance of Kelvin-Helmholtz instability is quantitatively captured by mixedness averaged align the direction of the pressure gradient. The role of compressibility and viscosity on mixing are investigated separately, both of which show two-stage effect. The underlying mechanism of the two-stage effect is interpreted as the development of large structures at the initial stage and the generation of small structures at the late stage. At the late stage, for a fixed time, a saturation phenomenon of viscosity is found that further increase of viscosity cannot see an evident decline in mixedness. The mixing statues of heavy and light fluids are not synchronous and the mixing of a RTI system is heterogenous. The results are helpful for understanding the mechanism of flow and mixing induced by RTI.
\end{abstract}

\maketitle

\section{Introduction}\label{sec:level1}
Rayleigh-Taylor instability (RTI) occurs at the perturbed interface of two fluids when the pressure gradient points from the heavy to the light fluid \citep{Rayleigh_1883,Taylor_1950,Lewis_1950}. RTI widely exists in both scientific and engineering scenarios, including supernova evolution in astrophysics \citep{Ribeyre_2004}, stratigraphic dynamics in geophysics \citep{Lev_2008}, and material mixing in chemical engineering \citep{Vinningland_2007}. The inertial confinement fusion (ICF), which is expected to be a viable approach to develop fusion energy, control of RTI is important for the successful ignition. If it allows free development of RTI, the applied compression is limited and the integrity of the fuel shell will be badly destroied, leading to the failure of ignition \citep{Lindl_2004,Edwards_2013,Wang_2016,Bradley_2014}. Although with the importance, the control of RTI is still a challenge due to unclearness of this strong nonlinear process, and better understanding of RTI will benefit both fundamental science and engineering practice \citep{Zhou_PR1_2017,Zhou_PR2_2017}.

Flow-induced by RTI is time-dependent. It undergoes different development stages, with both linear and nonlinear features, showing typical patterns of fingering and mixing of two fluids. In some cases, the flow can transit to a fully turbulent state. The RTI phenomenon was observed and upward to a scientific level by \citet{Rayleigh_1883}, and further analyzed by \citet{Taylor_1950}, \citet{Lewis_1950}, and et al. The concept of RTI was invented after then and a physical quantity of RTI, the growth rate was derived from physical modeling \citep{Kull_1991}. However, it is hard to generalize the whole process of RTI with a simple theory. Most theoretical works only focused on either the early linear (weakly non-linear) growing stage or the fully turbulent stage \citep{Goncharov_2002,Mikaelian_1998,Zhao_2018,Clark_2003}. Intervals between stages have yet not well understood \citep{Zhou_pof_2019}. Especially, the accompanied Kelvin-Helmholtz instability (KHI) induces large-scale vortices and mixing, which will enhance the non-linearity and make the instability even harder for the theoretical analysis. Besides theoretical analysis, experimental and numerical efforts was also made for better understanding RTI. In recent experiments, advanced apparatuses were employed, including planer laser-induced fluorescence to capture the RTI-induced mixing layer \citet{Waddell_2001,Wilkinson_2007} and high-speed video camera to visualize the abating effect of RTI on Richtmyer-Meshkov instability (RMI) by \citet{Luoxs_2018}. Although experiments provide good insight into the physics of RTI, the disadvantage is relatively expensive and time cost. The obtained information is limited by the ability of measuring techniques. Therefore, numerical simulation are widely adopted in RTI researches. Different methods in computational physics, including level-set method \citep{Chang_1996}, front tracking method \citep{Glimm_1986}, volume-of-fluid method \citep{Gerlach_2006}, smooth particle hydrodynamics \citep{Shadloo_2013}, large-eddy simulation \citep{Mellado_2005}, phase field method \citep{Celani_2009}, direct numerical simulation \citep{Mueschke_2009}, and etc., were already utilized for RTI simulation. With consideration of stabilizing and destabilizing effects on RTI, viscosity, compressibility, surface tension, and etc., were investigated in previous simulations \citep{Young_2001,Wei_2012,Liang_2019,Xie_2017,Hu_pof_2019}.

Generally, there are three kinds of fluid dynamic models in instability researches: the macroscopic model, mesoscopic model, and microscopic model. The macroscopic model, such as Euler and Navier-Stokes (NS), is based on the continuity hypothesis. According to the Chapman-Enskog theory, the NS includes only the first-order thermodynamic non-equilibrium (TNE) effect. This kind of model has been longly and mostly applied in fluid instability research. However, in complex fluid systems with mid-scale structure and material interfaces, the mean free path (or mean relaxation time) cannot be ignored compared to characteristic scale (or characteristic time) of the system, which means that discrete (or nonequilibrium ) effect becomes pronounced and consequently challenges the physical rationality of NS model. The microscopic models such as molecular dynamics simulation are the most precise method in capturing information for flows, but they are restricted by the computational cost in spatiotemporal scales. Recently, the mesoscopic discrete Boltzmann model (DBM) was proposed to deal with such mid-scale structure behaviors and is capable of simulation of a relatively wider spectrum of spatiotemporal scale. Via selecting appropriate kinetic moments to keep values, it can be designed according to the extent of TNE or discreteness which is aimed to investigate. The DBM describes the system from a wider research perspective. A DBM is equivalent to a traditional macroscopic fluid dynamic model plus a coarse-grained model describing TNE effects\citep{Xu_2018}. Currently, DBM has made significant progress in various hydrodynamic instabilities including RTI, RMI, and KHI, which brought a series of new insights of abundant TNE behaviors \citep{Xu_2018,Gan_2019,Lai_2016,Chen_2016,Chen_2018,Lin_2017,ZhangYD_2019}. Based on a multiple relaxation time DBM, besides investigating Prandtl number effects on RTI, \citet{Chen_2016,Chen_2018} uncovered the collaboration and competition roles between RMI and RTI and found high correlations between the nonuniformities of macroscopic quantity and nonequilibrium strength in the flow system. \citet{ZhangYD_2019} showed that via interfaces captured by appropriate TNE properties, the mixing of materials and that of internal energy resulted from hydrodynamic instability can be more conveniently probed. For example, the newly observed double-layer vortex has the correlation and difference of density mixing and temperature mixing/unification. \citet{Lai_2016} proposed a DBM model with acceleration term for RTI flow simulation and investigated the effect of compressibility on RTI, finding that the compressibility had a two-stage effect that delays RTI growing at the initial stage and accelerates it at the later stage. Meanwhile, they presented an interface-tracking method through TNE characteristics.

Along with RTI flow, material mixing is a focus due to many applications \citep{boffetta2020scaling}. The investigation of RTI mixing in complex flow systems requires a basic metric defining the mixing state. Several metrics were employed in previous researches. A coarse-grained indicator is the width of mixing layer defined as the distance between the amplitude of the spike and the bubble \citep{Cook_2002,Cabot_2006,Olson_2009,Dimonte_2004}, but the results and conclusions based on this metric may differ due to different definition of the edge of mixing layer \citep{ZhouY_pof_2019}. To obtain a fine-grained description of mixing, Zhou et al.\citep{ZhouY_pop_2019} proposed actual mixed mass which is decided by the local mass fraction of the two fluids and \citet{MaT_pop_2017} applied this metric in an advanced RTI experiment. Recently, \citet{Lin_2017} studied the mixing state through mixing entropy which was also determined by mass fraction of each fluid based on a two-component DBM.

Based on a previous work of discrete Boltzmann investigation of RTI made by \citet{Lai_2016}, this paper intends to deepen the study from two aspects, the development of fluids interface and late-stage mixing. Correspondently, we will seek an interface-tracking method through embedding tracers with DBM and study the process of mixing in RTI. One-way coupled tracers will be introduced to provide a synchronous Lagrangian viewpoint to the Eulerian flow field, and these tracers will open up a new angle for RTI analysis. The rest of this paper is organized as follows. The numerical method and physical modeling are presented in Section \S\ref{sec:method}. Simulation results are presented in Section \S\ref{sec:results} along with discussions on RTI flow and mixing considering the effects of compressibility and viscosity in Subsection \S\ref{subsec:compre} and Subsection \S\ref{subsec:visco} respectively. Finally, conclusions are made in Section \S\ref{sec:conclusions}.

\section{DBM for RTI flow coupled with tracers}\label{sec:method}

Regarding a complex flow system induced by RTI, though has been investigated for many years, it still faces two challenges. The first is how to model the process objectively. An accurate simulation is based on the effectiveness of the physical model and corresponding functions on the specific spatial and time scale. The error generated from physical modeling cannot be compensated for by any modification on algorithm. Secondly, it is a challenge to extract useful information from massive data generated from simulation. Only regarding these two challenges, it is possible to accurately model RTI flow and effectively obtain physics from the results.

\subsection{The construction of DBM} \label{DBM_cons}

DBM is a coarse-grained physical model based on non-equilibrium statistical physics. It selects a set of kinetic properties decided by corresponding kinetic moments to study the system. Specifically, the deeper the non-equilibrium is, the more kinetic moments of distribution function $f$ are needed. The initial several conserved kinetic moments(including density, momentum, and energy) are necessary to any non-equilibrium flows, whereas the non-conserved kinetic moments depend on the degree of non-equilibrium which is aimed to study. The Navier-Stokes equations are the case where the system deviates slightly from its thermodynamic equilibrium called the first-order non-equilibrium effect. In the case without external force, from the Boltzmann equation to a DBM, there are three steps: (i) simplification of the collision term, (ii) discretization of the particle velocity space, (iii) characterizing the nonequilibrium state and picking out meaningful nonequilibrium information. The first two steps are based on the properties of the system and seize the main contradiction. The third step is based on simulation results and seize the main contradiction. In the first two steps, the kinetic moments (or hydrodynamic quantities) used to describe the flow system must keep the same values before and after the simplification/discretization. In this way, the model is significantly simplified while the main features of described flow system are kept unchanged \citep{Xu_2018,Gan_2013}.

To obtain a DBM for RTI flow, the acceleration effect of gravity should be taken into account. A DBM with acceleration effect was proposed by \citet{Lai_2016},  through approximately replacing the distribution function $f$ in Boltzmann equation with its corresponding equilibrium distribution function $f^{eq}$. The simplified BGK-Boltzmann equation \citep{Bhatnagar_1954} with a force term \citep{HeShan_1998} is written as follow,
\begin{equation}
	\frac{\partial f}{\partial t} + \mathbf{v} \cdot \frac{\partial f}{\partial {\mathbf{r}}} - \frac{\mathbf{a}\cdot \left ( \mathbf{v} - \mathbf{u} \right )}{RT}f^{eq}=
	- \frac{1}{\tau} \left ( f - f^{eq} \right )
	\label{BGK-BE}
\end{equation}
where $f=f(\textbf{r}, \textbf{v}, t)$ is the distribution function of fluid molecules and $f^{eq}$ is the equilibrium distribution function. $\textbf{r}$ indicates spatial position, $\textbf{v}$ is the fluid velocity, $t$ indicates the time, $\textbf{a}$ corresponds to external body force, $R$ is the gas constant, $T$ is the temperature, and $\tau$ is the relaxation time.

Afterward, the continuum particle velocity space will be discretized. Through Chapman-Enskog expansion, a discrete Boltzmann model with at least 16 discrete velocities can cover fully two-dimensional compressible NSE on the description of fluid dynamics. According to previous researches, DBM provides credible result of flow simulation compared with classical solvers of NSE \citep{XuAG_2012,xu2015multiple,XuAG_2016,GanYB_2018}.

Another feature of DBM is that it provides a quantitative indicator of local thermodynamic nonequilibrium effects by defining nonequilibrium moments $\boldsymbol{\Delta}^{*}_{m, n}$ as \citep{Xu_2018,Gan_2013,LinCD_2016},
\begin{equation}
    \boldsymbol{\Delta}_{m,n}^{*}=\mathbf{M}_{m,n}^{*}\left ( f \right ) - \mathbf{M}_{m,n}^{*}\left ( f^{eq} \right )
    \label{eq02}
\end{equation}
where $\mathbf{M}_{m,n}^{*} \left( f \right)$ represents kinetic central moments of $f$, defined as $\int_{\mathbf{v}^*} \overbrace{\mathbf{v}^* \cdots \mathbf{v}^*}^{m} f d \mathbf{v}^* $ in continuum velocity space, in which used the relative velocity $\mathbf{v}^*$ as $\mathbf{v}^* = \mathbf{v} - \mathbf{u}$. $m$ indicates the number of velocity used in the moment and $n$ indicates the tensor order respectively. In discrete velocity space, it only needs to replace the integration and continuum velocity with summation and discrete velocity.

The moments introduced are interrelated and constitute a relatively complete description of the system's non-equilibrium states and behaviors. The collaboration of TNE quantities given by DBM presents a rough but quantitative indicator for the specific nonequilibrium state of a fluid system. Each independent component of the TNE quantities describes the nonequilibrium state from its own perspective. For example, $\boldsymbol{\Delta}_{2}^{*}$ indicates viscous stress tensor and $\boldsymbol{\Delta}_{3,1}^{*}$ indicates heat flux. $\boldsymbol{\Delta}_{3}^{*}$ and $\boldsymbol{\Delta}_{4,2}^{*}$ are higher-order non-equilibrium indicators beyond the traditional NS model and indicate the flux of viscous stress and of heat flux, respectively.

In order to initiate a RTI flow, initial and boundary conditions are set according to \citet{Lai_2016} and \citet{Scagliarini_2010}. The acceleration directs from top to bottom and the denser fluid locates above the lighter fluid. Initially at the interface, a sinusoidal disturbance occurs with the form $ \beta_{c}\left ( x \right )= \beta_{0} \cdot \cos \left(k x \right) $, where $\beta$ is the relative coordinate of the interface, $\beta_{0}$ is the amplitude of perturbation, and $k$ denotes the wavenumber related to wavelength as $k=\pi/\lambda$.

\subsection{Tracers coupled in RTI flow}

Tracers delineating fluid flow belong to a special kind of particles. Particles immersed in the fluid always have interaction with the surrounding phase but not necessarily have prominent influence on it. This is valid by assuming that the inertial effect of particles is extremely weak, and could be understood based on the indication of Stokes number ($Stk = u_0 \cdot t_0 / d_0 $, where $t_0$ is the dynamic relaxation time of particle, $u_0$ is local flow velocity, and $d_0$ is the diameter of particle). When $Stk \ll 1$, an immersed particle responds immediately to the surrounding change and moves closely according to local streamline. Therefore, the velocity of tracer is an indication of the surrounding fluid velocity. In order to obtain the tracer velocity from the surrounding fluid, the Dirac function is expected to be used according to \citet{peskin_2002},
\begin{equation}
	\mathbf{u}_p \left (\mathbf{r}_k \right )=\int_{\Omega} \mathbf{u} \left ( \mathbf{r} \right )
	\cdot \delta \left ( \left | \mathbf{r} - \mathbf{r}_k \right | \right ) d\Omega
	\label{velocity_delta}
\end{equation}
where $\mathbf{u}_p$ is the tracer velocity, and $\mathbf{r}_{k}$ indicates the spatial position of the $k$-th tracer. The subscript \textit{p} denotes tracers and \textit{k} indicate the tracer number. $\delta$ is the Dirac function. The above equation is universally satisfied through the flow field.

In the simulation, the flow field is usually discretized on Eulerian grids, so tracers may locate at the interval between grid nodes. Therefore, it is necessary to discretize continuum Dirac function through approximation,
\begin{equation}
	\mathbf{u}^t \left( \mathbf{r}_k \right )= \sum_{i,j} \mathbf{u}_{i,j}^{t} \psi \left( \mathbf{r}_{i,j},\mathbf{r}_k \right )
\end{equation}
where approximate discrete Dirac function $ \psi $ replaces the original continuum one. Two-dimensionally, $\psi$ equals to the product of two one-dimensional discrete Dirac functions $\varphi$,
\begin{equation}
	\psi \left( \mathbf{r}_{i,j},\mathbf{r}_k \right )=\psi \left( \left | \mathbf{r}_{i,j} - \mathbf{r}_k \right | \right)=\varphi \left ( \Delta r_x  \right ) \cdot \varphi \left ( \Delta r_y  \right )
\end{equation}

There are many choices for approximate discrete Dirac function, the one selected in this paper is a simplified form as \citet{peskin_2002}, which keeps high-order accuracy as well as easy to implement,
\begin{equation}
	\varphi \left ( \Delta r_x  \right )=\left \{ \begin{array} {ll}
		\left \{ 1+\textup{cos}\left [ \left ( \Delta r_x/\Delta x \right ) \cdot \pi/2 \right ] \right \},  & \Delta r_x \leq 2\Delta x \\
		0 ,& \Delta r_x >  2\Delta x \end{array} \right.
	\label{eq:varphifunc}
\end{equation}

\begin{figure}
	\centerline{\includegraphics[width=8cm]{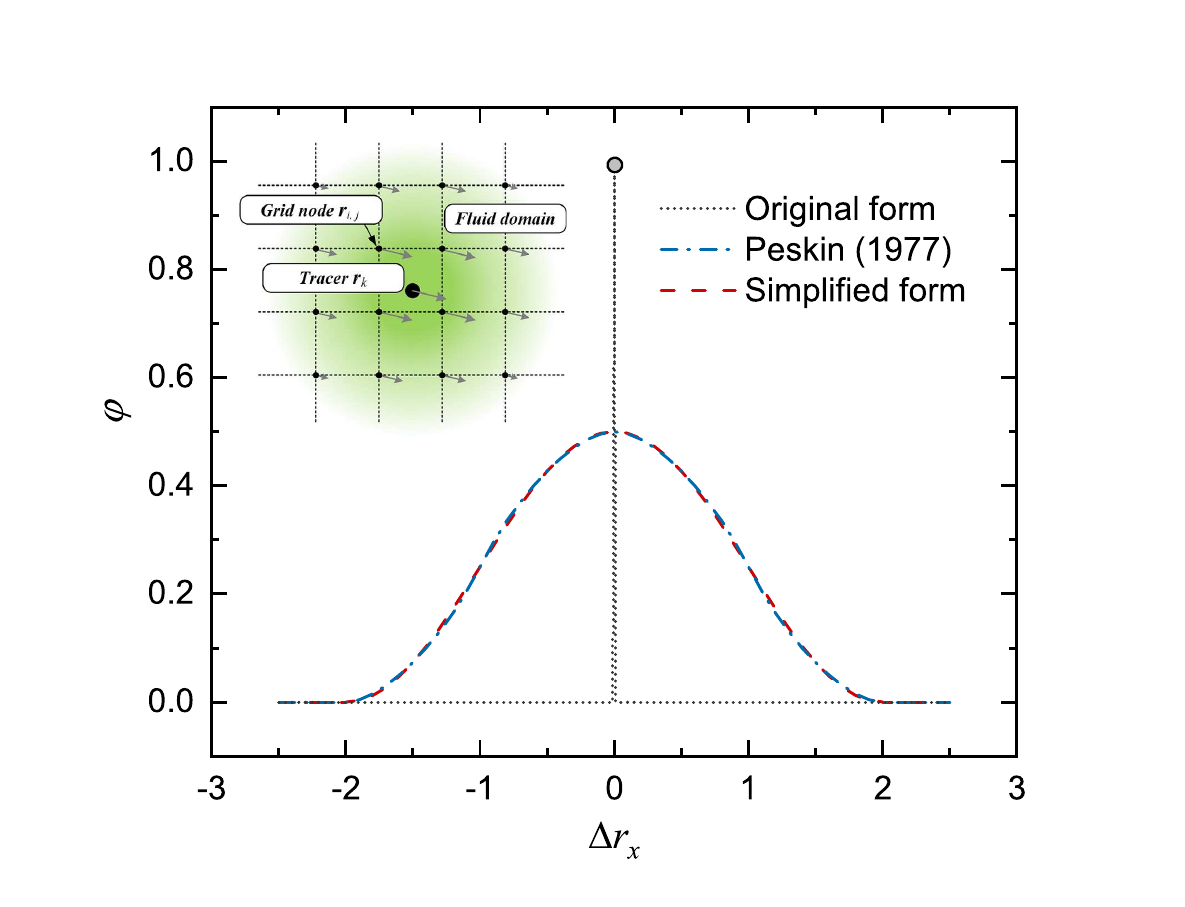}}
	\caption{Comparison of weights between the simplified form, the continuum form, and the one classical form. The schematic inner shows the implementation approach with the shade of different intensity indicating the weight of each nearby grids.}
	\label{fig:dirac_func}
\end{figure}

The schematic of implementing a discrete Dirac function on Eulerian grid and comparison of weights between the simplified form, the continuum form, and a classical form \citep{peskin_1977} is shown in Figure \ref{fig:dirac_func}.

After the velocity is determined, the trajectory of each tracer can be captured by motion equation as,
\begin{equation}
	\frac{\partial\mathbf{r}_k}{\partial t}=\mathbf{u}_p
	\left ( \mathbf{r}_k \right )
\end{equation}

To numerically capture the trajectory of each tracer, a 4th-order Runge-Kutta scheme is utilized for the discretization of the particle motion equation \citep{Ramsden_1991}. At each midpoint during Runge-Kutta processing, the tracer velocity is refreshed at the midpoint location, which guarantees a high-order accuracy of the calculation.

\section{Results and discussion}\label{sec:results}

A numerical simulation of 2D compressible RTI flow is conducted based on the above modeling and method. Mesh size is set of $1024 \times 256$ to guarantee enough resolution and to save computational cost. About 250000 tracer particles are randomly distributed to the whole region of simulation. Initial parameters are selected as Table \ref{tab:parameters} shows. Periodic boundary conditon and non-flux boundary condition are set to horizontal and vertical boundaries respectively for both fluids and tracers. The interface of the two fluids also separates tracers into two kinds. Tracers above are marked with \textit{type-a} and tracers below are marked with \textit{type-b}. This enables to distinguish the mixing of tracers during the simulation. Additionally, tracers set on the interface are marked as \textit{type-c} to record interfacial deformation and changes of physical quantities on the interface.

\begin{table}
	\begin{center}
		\def~{\hphantom{0}}
		\begin{tabular}{lccc}
			$Parameters$                      &   $Values$   \\[3pt]
			Time step ($dt$)                 &   2.0$\times$10$^{-5}$ \\
			Grid size ($dx$=$dy$)            &   1.0$\times$10$^{-3}$ \\
			Relaxation time ($\tau$)         &   5.0$\times$10$^{-5}$ \\
			Acceleration ($g$)               &   1.0 \\
			Upper fluid temperature ($T_u$)  &   1.0 \\
			Pressure at interface ($p_0$)       &   1.0 \\
			Atwood number ($At$)             &   0.6 \\
		\end{tabular}
		\caption{Initial physical parameters for RTI simulation}
		\label{tab:parameters}
	\end{center}
\end{table}

\subsection{Tracing RTI evolution}
\begin{figure*}
	\centerline{\includegraphics[width=16cm]{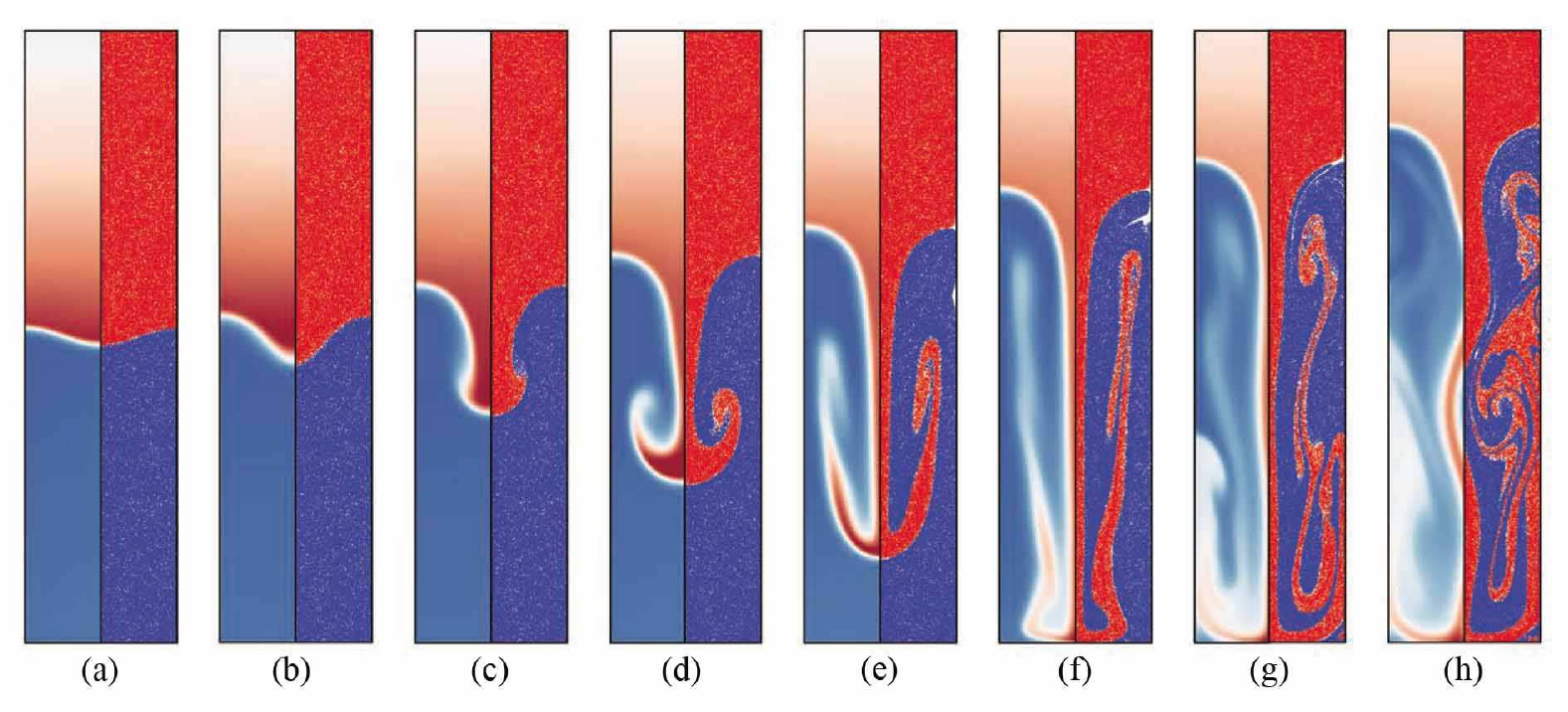}}
	\caption{Comparison between density contours and tracer distribution patterns during RTI development in a compressible miscible fluids system; the left half in each part shows density contours (the fluids interface is indistinct at the late time); the right half in each part shows tracer distribution (with clear evolution pattern at the late time). (a) $t = 0.1$; (b) $t = 0.5$; (c) $t = 1.0$; (d) $t =1.5$; (e) $t = 2.0$; (f) $t = 2.5$; (g) $t = 3.0$; (h) $t = 3.5$.}
	\label{fig:evl_pattern}
\end{figure*}

A comparison between evolution patterns of tracers and fluid density contours is shown in Figure \ref{fig:evl_pattern}. Fluid density is utilized for traditional evolution patterns of RTI. In an incompressible immiscible fluids system, the density contour is clear to show distinct difference between fluids in RTI. However, in the miscible and compressible fluids system which is investigated in this paper, the density cannot provide a clear interface due to transition layer of density is formed at the interface \citep{Gerya_2015,Kilkenny_1994,Yang_2002,Aziz_2000}. However, by marking two types of tracers with different colors, the tracer distribution provides a clear boundary of two fluids during each time of RTI evolution. From Figure \ref{fig:evl_pattern}(a) to Figure \ref{fig:evl_pattern}(d), the density (on the left half) and the tracer (on the right half) show similar evolution patterns of RTI. The similarity of two kinds of figures verifies the validity of results provided by tracers. With further development of RTI, the ability of tracers is demonstrated on the mixing description from Figure \ref{fig:evl_pattern}(e) to Figure \ref{fig:evl_pattern}(h). When fluids are mixed, it is hard to figure out the local component of two fluids. At this stage, the density cannot provide an exact delineation of mixing. Nevertheless, the distinct difference between two kinds of particles is kept during RTI mixing as shown in Figure \ref{fig:evl_pattern}(h). Clearly shown by the particle distribution patterns, fluids are mixed in the late stages and tiny vortex structures are generated. These tiny flow structures are important incentive to further mixing and their contribution will be analyzed in the following sections of this paper. Noted that, due to the discrete nature of tracers, there are some unavoidable white-noisy points on the tracer distribution pattern, but if enough tracers are included, the resolution will not interfere with the observation of the characteristic structures. Compared with a two-component DBM utilized to distinguish RTI mixing by \citet{Lin_2017}, which nearly doubles the computational cost of a single-component simulation, the current tracer method is more convenient and efficient.

\begin{figure*}
	\centering
	\subfigure{
		\begin{minipage}{4cm}
			\includegraphics[width=4cm]{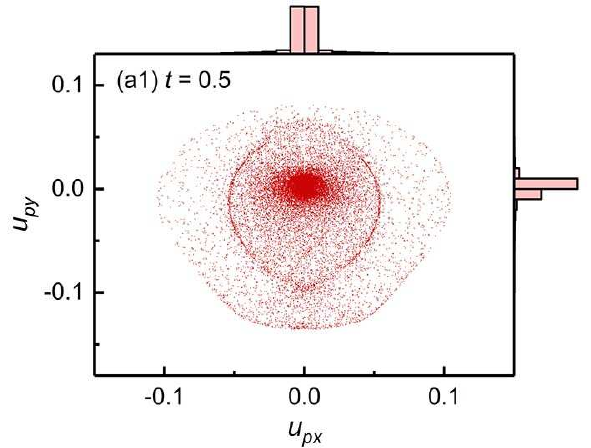}
		\end{minipage}
		\begin{minipage}{4cm}
			\includegraphics[width=4cm]{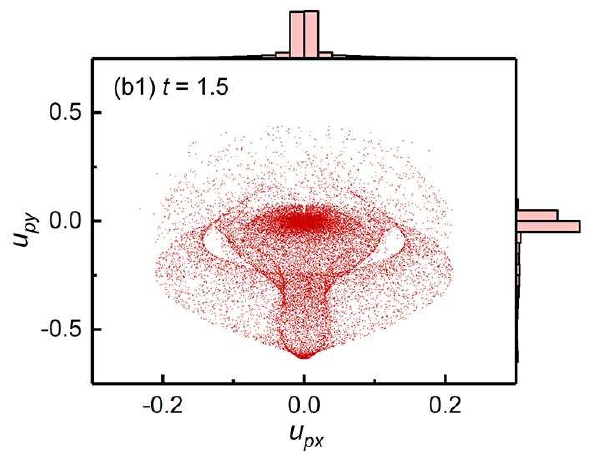}
		\end{minipage}
		\begin{minipage}{4cm}
			\includegraphics[width=4cm]{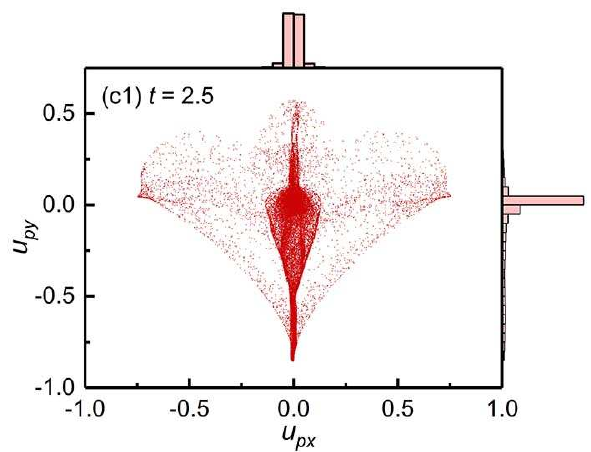}
		\end{minipage}
		\begin{minipage}{4cm}
			\includegraphics[width=4cm]{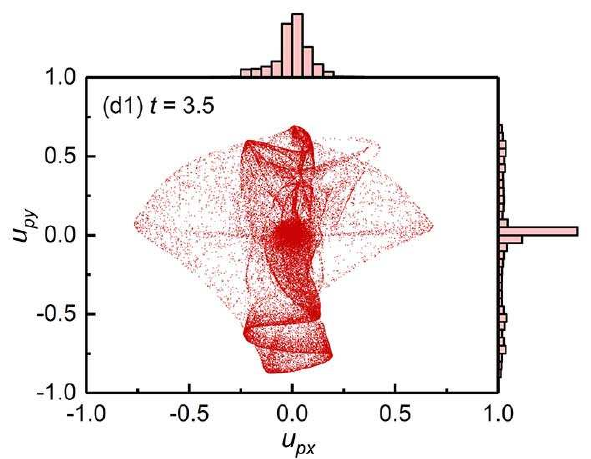}
		\end{minipage}
	}
	
	\subfigure{
		\begin{minipage}{4cm}
			\includegraphics[width=4cm]{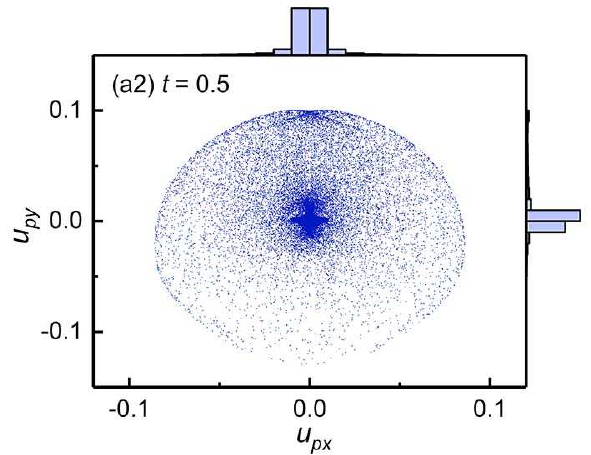}
		\end{minipage}
		\begin{minipage}{4cm}
			\includegraphics[width=4cm]{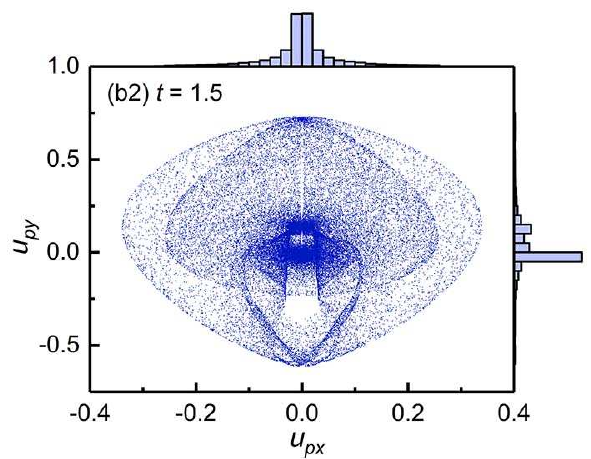}
		\end{minipage}
		
		\begin{minipage}{4cm}
			\includegraphics[width=4cm]{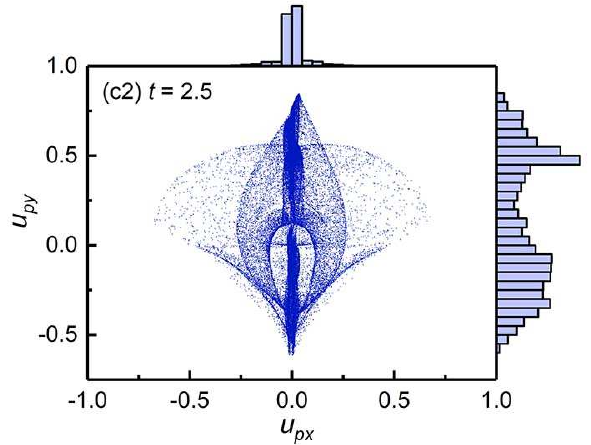}
		\end{minipage}
		\begin{minipage}{4cm}
			\includegraphics[width=4cm]{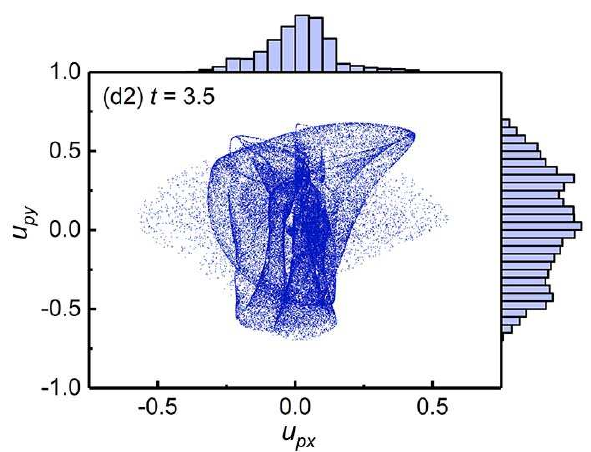}
		\end{minipage}
	}

	\caption{Velocity distribution of two kinds of tracers at four selected time, $t = 0.5, 1.5, 2.5, 3.5$, corresponding to delineated flows in Figure \ref{fig:evl_pattern}. The time selected roughly corresponds to the initiation of RTI, the initiation of KHI, late RTI growing, and late RTI mixing; (a1)-(d1) \textit{type-a} tracers in red color; (a2)-(d2) \textit{type-b} tracers in blue color. Histograms on the top and right side of each figure indicate the statistics of $u_{px}$ and $u_{py}$ respectively.}
	\label{fig:tracer_velocity_distribution}
\end{figure*}

Previously, most researches focused on the evolution of the interface of two fluids, a separated analysis on the heavy and light fluids is seldom. The velocity of each tracer represents local flow velocity and the two kinds of tracers may have different behavior corresponding to the characteristics of heavy and light fluids. A further investigation of the velocity distribution of tracers is meaningful for understanding and distinguishing stages of RTI flow. Therefore, a statistic of tracer velocity is presented and the tracer velocity distribution (TVD) is shown in Figure \ref{fig:tracer_velocity_distribution}. At $t = 0.5$, when the flow initiates, distribution of tracers concentrates at the original point where velocity equals zero, as shown by Figures \ref{fig:tracer_velocity_distribution}(a1) and (a2), indicating that the disturbance from the interface has not propagated long distance in the system. The outlines of two distribution patterns both form ring shapes. Along the outline, the tracer velocity is symmetrical to the original point with both positive and negative values. Initially, the movement of two fluids is consistent on the interface, with heavy fluid pushing the light fluid downward or light fluid boost heavy fluid upward at different places. Tracers distribution is not even along the outline. The lower part of the heavy fluid outline (Figure \ref{fig:tracer_velocity_distribution}(a1)) and the upper part of the light fluid outline (Figure \ref{fig:tracer_velocity_distribution}(a2)) is denser. During the movement, both heavy and light fluid encounters the hindrance from the other, leading to the compression of fluids at the interface. Gradually, the symmetry between the upper and lower halves of the two kinds of TVD patterns fully diminishes. At $t = 1.5$, a layered pattern is formed in both TVD, as shown by Figures \ref{fig:tracer_velocity_distribution}(b1) and \ref{fig:tracer_velocity_distribution}(b2). The inner structures of two TVDs are different. As shown in Figure \ref{fig:tracer_velocity_distribution}(b1), a `drip' develops from the origin and encircled by a 'crown', indicating that a small part of \textit{type-a} tracers aligned with the spike breaks through the hindrance of surrounding light fluid during RTI. At the same time, two inside structures are formed vertically inside an 'envelope', see Figure \ref{fig:tracer_velocity_distribution}(b2). As shown by histograms in Figures \ref{fig:tracer_velocity_distribution}(a1), \ref{fig:tracer_velocity_distribution}(a2), \ref{fig:tracer_velocity_distribution}(b1), and \ref{fig:tracer_velocity_distribution}(b2), both horizontal velocity ($u_{px}$) and vertical velocity ($u_{py}$) of \textit{type-a} and \textit{type-b} tracers distribute normally although with small deviation. At $t = 2.5$, two petunia-like patterns are shown in Figures \ref{fig:tracer_velocity_distribution}(c1) and \ref{fig:tracer_velocity_distribution}(c2). The outlines of two TVD patterns resemble petals with a long `stamen' in the center, indicating that many tracers are with zero horizontal velocity but the vertical velocity of tracers spans for a large range. The spike maintains its flow direction when it penetrates the light fluid, so the `stamen' in Figure \ref{fig:tracer_velocity_distribution}(c1) is slender than that in Figure \ref{fig:tracer_velocity_distribution}(c2). At $t = 2.5$, the distribution of $u_{px}$ of both \textit{type-a} and \textit{type-b} tracers centralize at original points and are close to normal distribution. However, a distinct double-peak distribution is observed from $u_{py}$ histogram of \textit{type-b} tracers. At the time, KHI develops and leads to the vortex motion of both heavy and light fluids. Two fluids roll together in the same vortex. A small swelling of \textit{type-a} $u_{py}$ histogram indicates that the KHI vortex consists of a small portion of heavy fluids and the double-peak in \textit{type-b} $u_{py}$ histogram shows that the light fluid constitutes mainly in KHI vortex. At $t = 3.5$, when there is no distinct spike and bubble in the RTI system, spiral patterns are observed in two TVDs, as shown by Figures \ref{fig:tracer_velocity_distribution}(d1) and \ref{fig:tracer_velocity_distribution}(d2). The $u_{px}$ histograms of both the \textit{type-a} and \textit{type-b} tracers cover a longer range, so tracers transport randomly in this stage without evident groups. The two peaks in $u_{py}$ histogram of \textit{type-b} tracers disappear, and approximately normal distribution is formed. Distinct difference is observed between $u_{py}$ histogram of \textit{type-a} and \textit{type-b} tracers. It concludes that the mixing statues of heavy and light fluids are not synchronous and the mixing of RTI system is heterogeneous. Here presents the most interesting observation of TVDs, for further understanding, investigation of the dynamic process is still important to be made.

\subsection{Interfacial tracking manifestation}

\begin{figure}
	\centerline{\includegraphics[width=8cm]{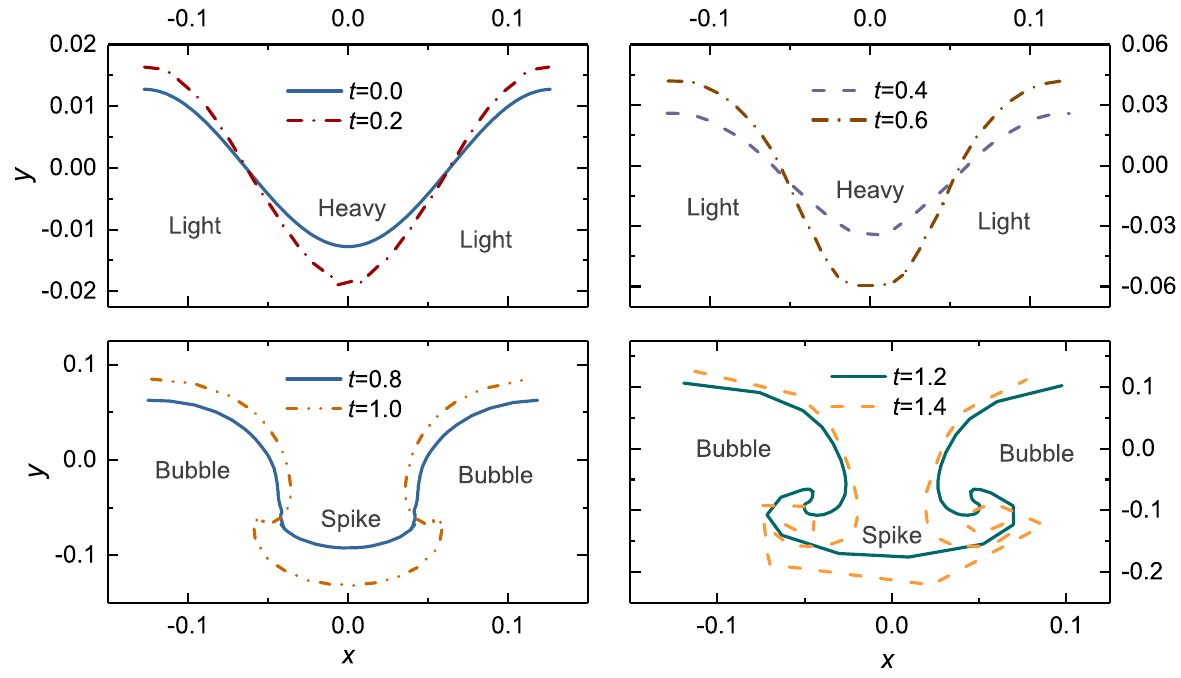}}
	\caption{Deformation of the interface during RTI growing illustrated by tracers.}
	\label{fig:04}
\end{figure}

The interface tracking technique presents a problem in complex fluid systems and the tracer method brings an effective measure in interface tracking. Initially, the number of 128 \textit{type-c} particles are distributed at the intermittent interface. Then, through recording the positions of \textit{type-c} particles in RTI evolution, we can easily delineate characteristics of the interface during RTI growing. Examples are shown in Figure \ref{fig:04}.

Initially, a start-up slight perturbation exists at the interface. At $t = 0.2$, fingers of two fluids (spike and bubble) are formed at the interface and then penetrate each other. However, the developing rates of the two parts are not consistent. At $t = 0.2$, the spike crosses the position of $y = -0.015$ while the bubble just reaches the position of $y = 0.015$. Then, the spike grows still faster than the bubble, which is indicated by figures of late times as $t = 0.4$, 0.6, and 0.8. Due to acceleration, the interface continually deforms, while at $t = 1.0$, the velocity difference on the shear direction of the interface initiates observable KHI, with two large vortices taking their shapes at each side of the spike, which is in accordance with the classical observation made by \citet{Kull_1991}. The development of KHI is a critical step toward the opening of turbulent mixing \citep{Peltier_2003} and it also accounts for the formation of mushroom-shaped flow structure during the RTI development \citep{Zabusky_1999}. Because the interface is stretched by further development of KHI, the limited number of tracers distribute more sparsely at a late time, so it is impossible to sketch the profile of the interface precisely as a decreased number of tracers locate at the front, shown by the interface profile at $t = 1.4$. After that, the tracer description for the interface is unacceptable. Therefore, it is important to design a tracking approach with appropriate number or concentration of tracer afore the RTI interfacial tracking.

Furthermore, as the main place for hydro-instability initiation and development, it is worthwhile to pay special attention to physical features of the interface, which may also be helpful for controlling instability in ICF \citep{Betti_1998, YuCX_2018}. Also, the design and improvement of all the interface tracking techniques need a better physical understanding of the interface. Fundamentally, thermodynamic nonequilibrium (TNE) is the driving force of fluid system evolution. The complexity of non-equilibrium behavior determines the diversity of the description of non-equilibrium degree. Besides, the TNE behaviors of the interface are indispensable content in the RTI system.

Fundamentally, Knudsen number, viscosity, heat conductivity, the macroscopic gradients of density, flow velocity, temperature, and pressure are commonly used to characterize the degree of non-equilibrium. They all describe the degree of nonequilibrium of the system from their respective perspectives. But they are also highly condensed and averaged coarse-grained descriptions of certain information. Much specific information about non-equilibrium states, the specific allocation of internal energy in different degrees of freedom, etc., through which they are invisible and cannot be directly studied. Therefore, in addition to the description mentioned above, we need a more detailed description of TNE. In this respect, the non-conservative kinetic moments of $(f - f^{eq})$ (as introduced in Section \ref{DBM_cons}) can be used for a more detailed description of the nonequilibrium behavior\citep{Xu_2018}.

By extracting the TNE information of \textit{type-c} tracers through interpolating the TNE information of fluid field, we hence demonstrate the TNE property of interface. In the following analysis of interfacial TNE, the period of $t$ = 0 to 1.2 is selected during which the application of \textit{type-c} tracers is acceptable. The mean TNE of interface is then calculated as Equation \eqref{eq:intpo},

\begin{equation}
\overline{\boldsymbol{\Delta}}_{m,n}^{*}=\frac {1}{\Lambda_{i}} \int_{\Lambda_{i}} \boldsymbol{\Delta}_{m,n}^{*} \left (x,y \right ) dl \approx \frac
{\sum_{1}^{N} \boldsymbol{\Delta}_{m,n,k}^{*} }{N}
\label{eq:intpo}
\end{equation}
where $\Lambda_{i}$ represents the length of interface and $N$ is the total number of \textit{type-c} tracers.

\begin{figure*}
	\centering
	\subfigure{
		\begin{minipage}{6.5cm}
			\includegraphics[width=6cm]{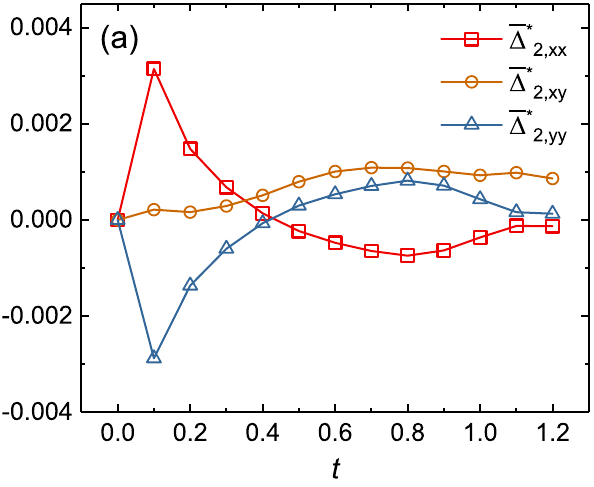}
		\end{minipage}
		\begin{minipage}{6.5cm}
			\includegraphics[width=6cm]{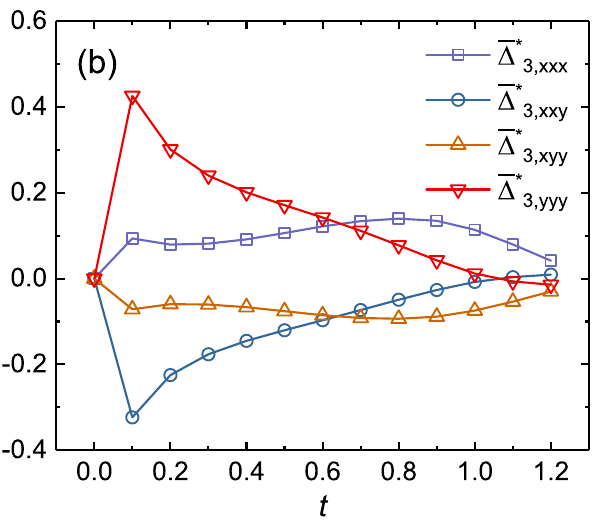}
		\end{minipage}
	}
	
	\subfigure{
		\begin{minipage}{6.5cm}
			\includegraphics[width=6cm]{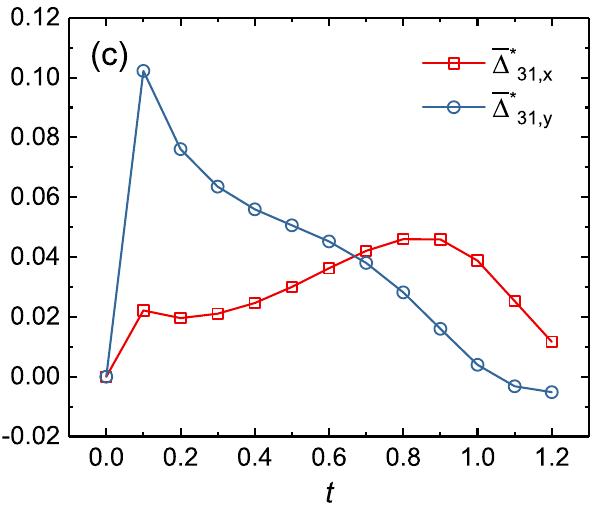}
		\end{minipage}
		\begin{minipage}{6.5cm}
			\includegraphics[width=6cm]{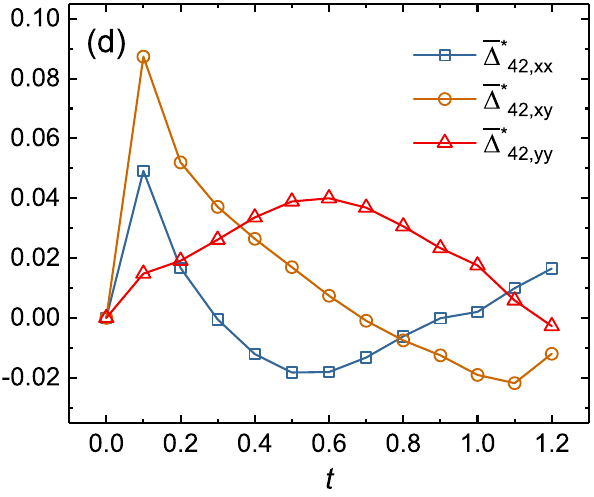}
		\end{minipage}
	}
	
	\caption{Averages of different TNE components on the interface versus time. (a) $\overline{\boldsymbol{\Delta}}^{*}_{2}$; (b) $\overline{\boldsymbol{\Delta}}^{*}_{3}$; (c) $\overline{\boldsymbol{\Delta}}^{*}_{3,1}$; (d) $\overline{\boldsymbol{\Delta}}^{*}_{4,2}$.}
	\label{fig:TNE_components}
\end{figure*}

From the averages over the interface of each TNE component ($\overline{\boldsymbol{\Delta}}^{*}_{2}$, $\overline{\boldsymbol{\Delta}}^{*}_{3}$, $\overline{\boldsymbol{\Delta}}^{*}_{3,1}$, and $\overline{\boldsymbol{\Delta}}^{*}_{4,2}$) in Figures \ref{fig:TNE_components}(a)-(d), we observe that most curves have a peak value at bout $t = 0.1$. At this time, there is no obvious distortion at the interface, which means the driving potential is accumulated firstly and then activates the development of instability. Figure \ref{fig:TNE_components}(a) shows the changes of $\overline{\Delta}^{*}_{2,xx}$, $\overline{\Delta}^{*}_{2,xy}$, and $\overline{\Delta}^{*}_{2,yy}$ over time, which corresponds to viscous stress in dimension. During the development, curves of $\overline{\Delta}^{*}_{2,xx}$ and $\overline{\Delta}^{*}_{2,yy}$ are nearly symmetrical along TNE value of 0, indicating the homogeneous of normal pressure on the interface. During the development, $\overline{\Delta}^{*}_{2,xy}$ fluctuates and gradually increases, indicating that the momentum flux exchange on the shear direction is slightly enhanced during interface deformation. In Figure \ref{fig:TNE_components}(b), the changes of four independent components of $\overline{\boldsymbol{\Delta}}^{*}_{3}$ ($\overline{\Delta}^{*}_{3,xxx}$, $\overline{\Delta}^{*}_{3,xxy}$, $\overline{\Delta}^{*}_{3,xyy}$, and $\overline{\Delta}^{*}_{3,yyy}$) are shown with time. $\overline{\Delta}^{*}_{3,xxx}$ and $\overline{\Delta}^{*}_{3,xxy}$ indicate the flux of $x$ component of kinetic energy on $x$ and $y$ direction respectively. They deviate from equilibrium state in the opposite direction. Similarly, $\overline{\Delta}^{*}_{3,xyy}$ and $\overline{\Delta}^{*}_{3,yyy}$ indicate the flux of $y$ component of kinetic energy on $x$ and $y$ direction respectively. They deviate from the equilibrium state in the opposite direction either. Good symmetry is found between $\overline{\Delta}^{*}_{3,xxx}$ and $\overline{\Delta}^{*}_{3,xyy}$ along TNE value of 0. However, curves of $\overline{\Delta}^{*}_{3,xxy}$ and $\overline{\Delta}^{*}_{3,yyy}$ are not well symmetrical. Specifically, the peak value of $\overline{\Delta}^{*}_{3,yyy}$ is greater than that of $\overline{\Delta}^{*}_{3,xxy}$. The value of $\overline{\Delta}^{*}_{3,xxy}$ and $\overline{\Delta}^{*}_{3,yyy}$ both decrease over time, but $\overline{\Delta}^{*}_{3,xxx}$ and $\overline{\Delta}^{*}_{3,xyy}$ increase firstly and then decline at about $t = 0.8$. According to Figure \ref{fig:TNE_components}(c), $\overline{\Delta}^{*}_{31,y}$ reaches its peak value at $t = 0.1$ and monotonously drops then. The $\overline{\Delta}^{*}_{31,y}$ curve experiences a slight decline near $t = 0.1$ and rises again and then drops. Initially at the interface, the energy flux indicated by $\overline{\boldsymbol{\Delta}}^{*}_{3,1}$ has larger value on vertical direction ($\overline{\Delta}^{*}_{31,y}$) than horizontal direction ($\overline{\Delta}^{*}_{31,x}$), but the $\overline{\Delta}^{*}_{31,x}$ surpasses $\overline{\Delta}^{*}_{31,y}$ at about $t = 0.67$, delineating that the vertical energy flux is not dominant over the horizontal flux in the system. From dimensional analysis, $\boldsymbol{\Delta}^{*}_{4,2}$ corresponds to the flux of energy flux, which indicates a more complex physical process. It is found that two components of $\overline{\boldsymbol{\Delta}}^{*}_{4,2}$, $\overline{\Delta}^{*}_{42,xx}$ and $\overline{\Delta}^{*}_{42,xy}$, reach their peak values at $t = 0.1$, and decline until $t = 0.5$ and 1.1, respectively. Also, both the two components rise during the following time.

In the phase space constituted by each independent component of TNE provided by DBM, any points in this phase space correspond to a specific TNE state. The distance from a specific state to the original point in the non-equilibrium phase space defines the extent of non-equilibrium roughly, see Equation \eqref{eq:mean_TNE}.
\begin{equation}
	D_{\Lambda}=\sqrt{ \left( \boldsymbol{\Delta}_{2}^{*} \right)^2 + \left ( \boldsymbol{\Delta}_{3}^{*} \right)^2 + \left( \boldsymbol{\Delta}_{3,1}^{*} \right)^2 + \left( \boldsymbol{\Delta}_{4,2}^{*} \right)^2 }
	\label{eq:mean_TNE}
\end{equation}
where $D_{\Lambda}$ indicates interfacial TNE strength by measuring the distance between a specific point in phase space and the origin point. $\boldsymbol{\Delta}_{2}^{*}$, $\boldsymbol{\Delta}_{3}^{*}$, $\boldsymbol{\Delta}_{3,1}^{*}$, and $\boldsymbol{\Delta}_{4,2}^{*}$ are different components of a TNE state.

\begin{figure}
	\centerline{\includegraphics[width=8cm]{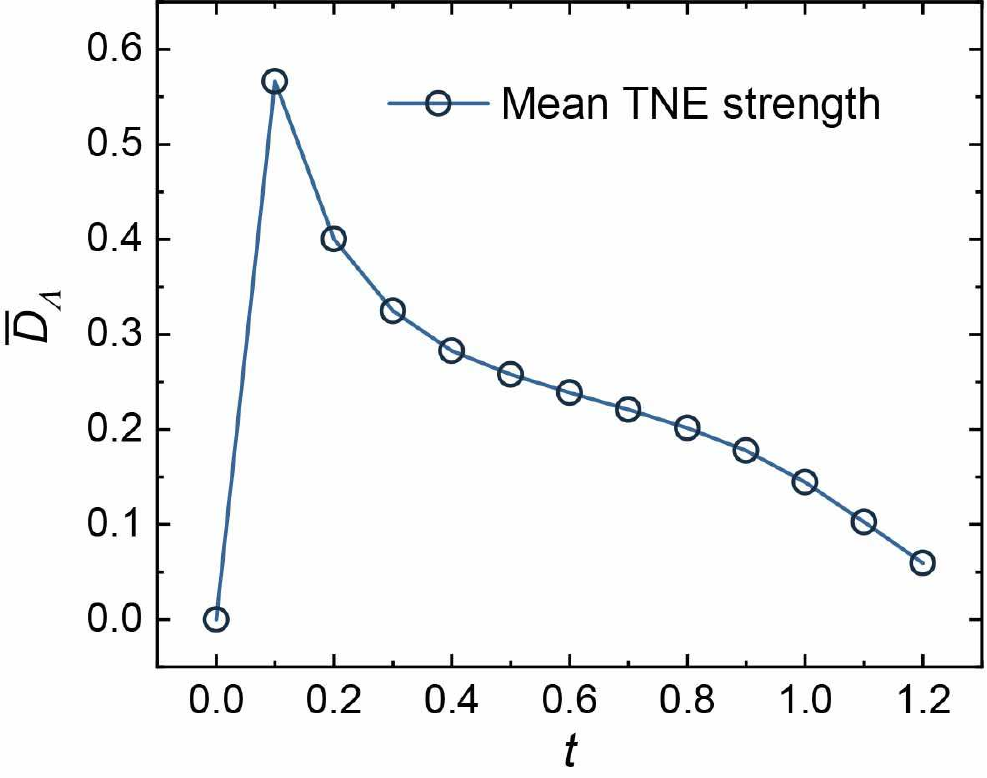}}
	\caption{Changes of mean TNE strength on the interface over time.}
	\label{fig:mean_TNE}
\end{figure}

Figure \ref{fig:mean_TNE} shows the evolution of average TNE strength on interface ($\overline{D}_{\Lambda}$). The mean TNE strength has its maximum value at the initial stage as $t = 0.1$, and then gradually decreases. The reason is that TNE triggers the initial instability of the system at the start of the simulation, and then the system evolves spontaneously to a new equilibrium state, so there is a peak of the mean TNE strength at the beginning and gradual decreases subsequently.

The analysis is made out of our most present knowledge, but TNE phenomena are so fruitful that cannot be fully investigated, so we leave it for a future study. In sum, a specific non-equilibrium state requires a comprehensive consideration of various non-equilibrium components, and their collaboration is combined to give a more complete description of the flow system.

\subsection{Delineation of RTI mixing}
As well as delineation of complex flows, the tracers provide an approach for analyzing the mixing state of fluids. In order to quantify the extent of mixing, the local mixedness ($\chi_p$) is defined based on the spatial information of tracers as,
\begin{equation}
    \chi_p \left (x,y \right )=4 \frac {c_a \left (x,y \right )\cdot c_b\left (x,y \right )}{\left [ c_a \left (x,y \right )+ c_b\left (x,y \right ) \right ]^2}
    \label{eq:local_mixedness}
\end{equation}
where $c_a$ and $c_b$ are local number density of \textit{type-a} and \textit{type-b} tracers respectively. When there is no \textit{type-a} or \textit{type-b} tracers locally , $\chi_p$ equals to the minimum value 0, while if $c_a = c_b$, $\chi_p$ equals to the maximum value 1.0. Thus, $\chi_p$ ranges from 0 to 1.0, indicating from the none mixing to the fully mixing state.

To evaluate the mixing state of the whole field, the averaged mixedness ($\overline{\chi}_p$) is calculated according to Equation \eqref{eq:avm},
\begin{equation}
\overline{\chi}_p=\frac{1}{\Omega}\int_{\Omega} \chi_p \left (x,y \right ) d \Omega
\label{eq:avm}
\end{equation}
where $\Omega$ represents the whole field.

\begin{figure}
    \centerline{\includegraphics[width=8cm]{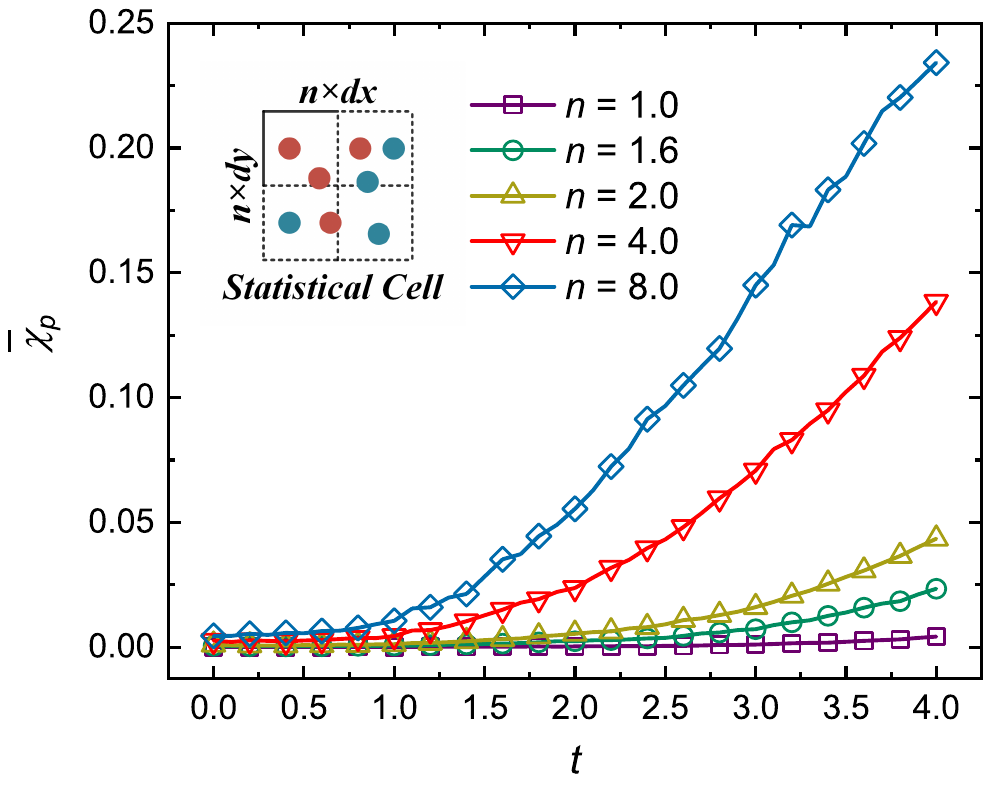}}
    \caption{Growth of average mixedness versus time (based on different statistical cells).}
    \label{fig:DifferentSCs}
\end{figure}

Transferring information of discrete tracers into continuous number density ($c_a$ or $c_b$) is a key step. By dividing the simulation area into a limited number of statistical cells, we could calculate the number density of tracers in a cell. However, the cell should be selected with proper size in order to make an accurate demonstration of mixing ($ds_x$ and $ds_y$ are cell lengths on $x$ and $y$ directions respectively). Statistical cells with sizes of $n$ times of the grid size ($ds_x = n \cdot dx$ and $ds_y = n \cdot dy$) is employed, and $n$ ranges from 1.0 to 8.0 is presented. The growth of $\overline{\chi}_p$ over time under different statistical cells is shown in Figure \ref{fig:DifferentSCs}. All the curves show similar growing process of mixedness. Initially, the mixedness is nearly zero, while after a short period, the mixing accelerates and mixedness grows rapidly. Shown by results from different statistical cells that, the mixedness based on small cells ($n = 1.0$ for example) is relatively low during the whole growing process, and mixedness based on large cells ($n = 8.0$ for example) shows evident perturbation. A smooth growing curve can be achieved based on medium-sized statistical cells. In this case, we select $n = 2.0$ mainly for the following research.

\begin{figure}
    \includegraphics[width=7cm]{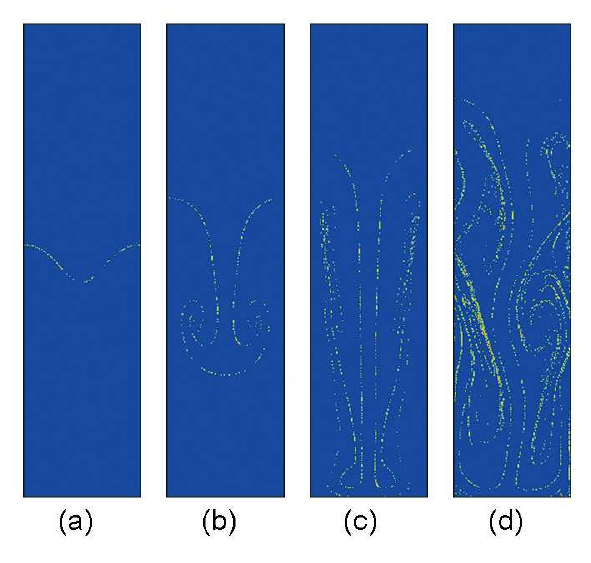}
    \caption{Mixedness distribution based on statistical cell with size of $n = 2.0$. (a)-(d) correspond to four selected time, $t = 0.5$, 1.5, 2.5, and 3.5.}
    \label{fig:mixedness_distri}
\end{figure}

Figure \ref{fig:mixedness_distri} shows the distribution of mixedness based on $n = 2.0$. Four representative time ($t = 0.5$, 1.5, 2.5, and 3.5) of RTI development are selected. It is found that the mixing mainly occurs at the contacting area of two fluids. As the interface being elongated and distorted in RTI flow, the contacting area increases, leading to further mixing.

Through integrating the mixedness ($\chi_p \left (x,y \right )$) along $y$ direction and then dividing the height of simulation region $L_y$, the average mixedness in vertical direction ($\widetilde{\chi}_p$) is,
\begin{equation}
    \widetilde{\chi}_p \left(x \right )= \frac{1}{L_y} \int_{L_y} \chi_p \left (x,y \right ) ds_y
    \label{eq:vm}
\end{equation}

\begin{figure}
    \centerline{\includegraphics[width=9cm]{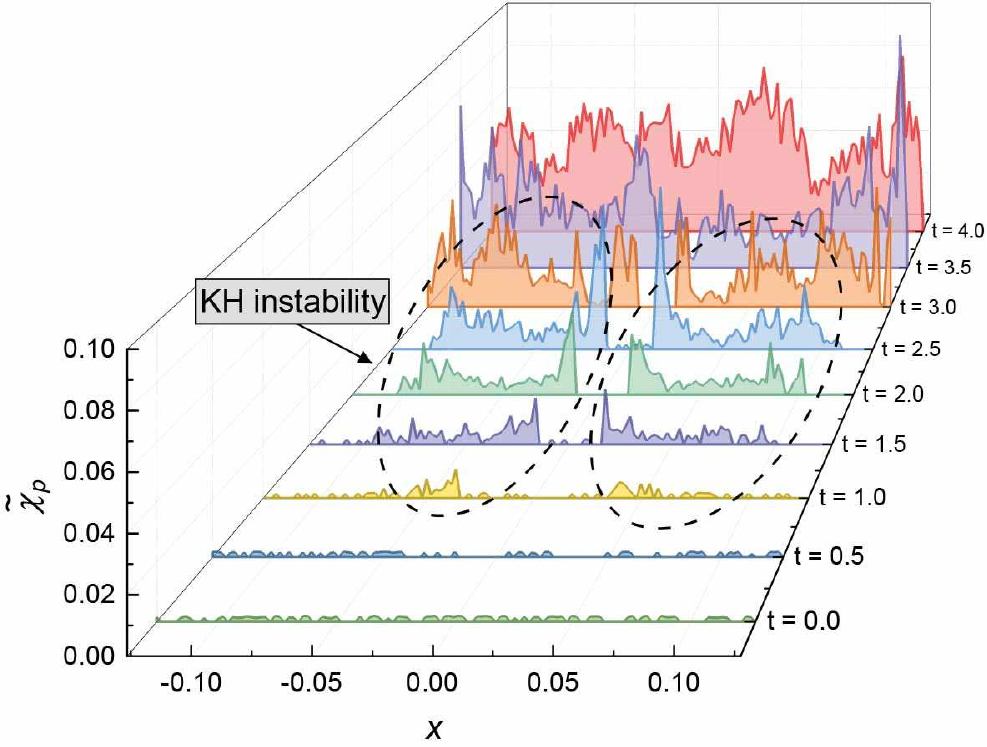}}
    \caption{Profiles of average mixedness $\widetilde{\chi}_p$ along the horizontal direction at different times. Circles emphasize the mixing region with Kelvin-Helmholtz (KH) instability.}
    \label{fig:vm}
\end{figure}

The evolution pattern of $\widetilde{\chi}_p$ based on $n = 2.0$ is shown in Figure \ref{fig:vm}. Initially, $\widetilde{\chi}_p$ is almost zero and fluctuates along $x$ axis. As RTI develops, two peaks of $\widetilde{\chi}_p$ appear along $x$ axis. Comparing the appearing time and positions of the two mixedness peaks with RTI flow patterns in Figure \ref{fig:evl_pattern}, the two peaks coincide with the development of KHI, which occurs at the interface of the two fluids on both sides. The two peaks of the $\widetilde{\chi}_p$ gradually increase with time, indicating the enhancement of KHI over time. Finally, as $\widetilde{\chi}_p$ is almost even, KHI vortex disappears and the mixing is found on smaller scales.

The evolution of $\widetilde{\chi}_p$ inspires us that it provides a good indicator for KHI occurrence, with both the time and spatial range of KHI. Values of $\widetilde{\chi}_p$ also indicate the intensity of KHI during the development. $\widetilde{\chi}_p$ presents at least two advantages on capturing KHI. There is no need of visualizing the flow and manually differentiating the rotation region. The calculation of $\widetilde{\chi}_p$ is done separately from the main simulation, which means it is a post-process and can be done at any time with simulation data.

Additionally, the distribution of $\widetilde{\chi}_p$ based on two different statistical cells with size of $n = 1.0$ and $n = 8.0$ are provided by Figures \ref{fig:a1}(a) and \ref{fig:a1}(b) in appendix \ref{appA}. The two figures show similar results with Figure \ref{fig:vm} and the details are discussed.

\subsection{Compressibility effect on RTI mixing}\label{subsec:compre}
Compressibility plays an important role in the growth of RTI as pointed by many researchers \citep{Bernstein_1983, Kull_1991, Gauthier_2010}. However, the effect of compressibility on RTI mixing has not been well analyzed. In this case, we intend to discuss how compressibility impacts RTI mixing. Firstly, a compression factor is defined through dimensional analysis \citep{Lai_2016, jin_2005RTI}.

The speed of sound ($c_s$) in ideal gas hypothesis is given as,
\begin{equation}
c_s = \sqrt{\frac{\gamma p}{\rho}}
\label{eq:idgas}
\end{equation}
where $\gamma$ is the isentropic expansion factor.

The Navier-Stokes equation coupled with ideal gas equation ($p = \rho R T$) then writes,
\begin{equation}
\frac{\partial \mathbf{u}}{\partial t} +
\mathbf{u} \cdot \nabla \mathbf{u} +
\frac{c_{s}^{2}}{\gamma \rho} \nabla \rho = \mathbf{g} +
\frac{\nabla \cdot \boldsymbol{\sigma}}{\rho}
\label{eq:d1}
\end{equation}
where $\boldsymbol{\sigma}$ is the stress tensor expressed as $\boldsymbol{\sigma} = \mu \left[\nabla \mathbf{u} + \left( \nabla \mathbf{u} \right)^{T} - \frac{2}{3} \left(\nabla \cdot \mathbf{u} \right) \boldsymbol{I} \right]$.

The sound speed $c_s$ and relaxation time $\tau$ are kept in the equation. The wavenumber $k$ and the acceleration $g$ are selected as basic units. Therefore, the typical length is $1/k$, the typical time is $\left (g \cdot k \right)^{-1/2}$, and the typical velocity is $\left (g/k \right)^{1/2}$. Navier-Stokes equation is transferred to a dimensionless form,
\begin{equation}
\frac{\partial \mathbf{u}^{'}}{\partial t^{'}}+\mathbf{u}^{'} \cdot \nabla^{'} \mathbf{u}^{'}+\frac{1}{H_c} \frac {\nabla^{'} \rho}{\rho} = 1 + \frac{\nabla^{'} \cdot \boldsymbol{\sigma}^{'}}{\rho}
\label{eq:NSND}
\end{equation}
where the viscous term at the end of the equation writes,
\begin{equation}
\frac{\nabla^{'} \cdot \boldsymbol{\sigma}^{'}}{\rho} =
\frac{H_{\tau}}{\gamma H_c} \nabla^{'} \cdot \left[\nabla^{'} \mathbf{u}^{'} + \left( \nabla^{'} \mathbf{u}^{'} \right)^{T} - \frac{2}{3} \left(\nabla^{'} \cdot \mathbf{u}^{'} \right) \boldsymbol{I} \right]
\label{eq:STND}
\end{equation}

In the above equation, $H_c$ is the square of the ratio of the typical speed to the sound speed, defined as the compression factor,
\begin{equation}
H_c=\left(\frac{\sqrt{g/k}}{c_s} \right)^2
\label{eq:cf}
\end{equation}

$H_{\tau}$ is the ratio of relaxation time to typical time,
\begin{equation}
H_\tau=\frac{\tau}{\left (g \cdot k \right)^{-1/2}}
\label{eq:Htau}
\end{equation}

It is noted that following the same dimensional analysis, BGK-Boltzmann equation can be transferred as,
\begin{equation}
\frac{\partial f^{'}}{\partial t^{'}}+\mathbf{v}^{'} \frac {\partial f^{'}}{\partial \mathbf{r}^{'}} + \frac{\partial f^{'}}{\partial \mathbf{v}^{'}} = - \frac{1}{H_\tau} \left( f^{'} - f^{eq'} \right)
\label{eq:NDBGKB}
\end{equation}
where $H_{\tau}$ is in accord with the physical meaning of the Knudsen number, indicating the nonequilibrium of the flow system, so $H_{\tau}$ is defined as Knudsen number.

In the following analysis on compressibility effect, the sound speed in upper fluid is chosen as $c_s$. At the same time, in order to keep the nonequilibrium scale of different compressible systems unchange, the Knudsen number ($H_\tau$) is kept constant as an initial condition, which is accomplished by adjusting different $\tau$ and $g$. $H_\tau = 2.483 \times 10^{-4}$ and six different characteristic compression factors $H_c = 0.011596$, 0.017393, 0.023191, 0.028989, 0.034787, and 0.040584, are selected. In the discussion, physical time is calculated according to a typical time $t^{*} = t/\left [2\pi/ \left( g \cdot k \right) \right]^{-1/2}$, which is derived from initial perturbation on interface.

\begin{figure*}
    \centering
    \begin{minipage}{8cm}
        \includegraphics[width=8cm]{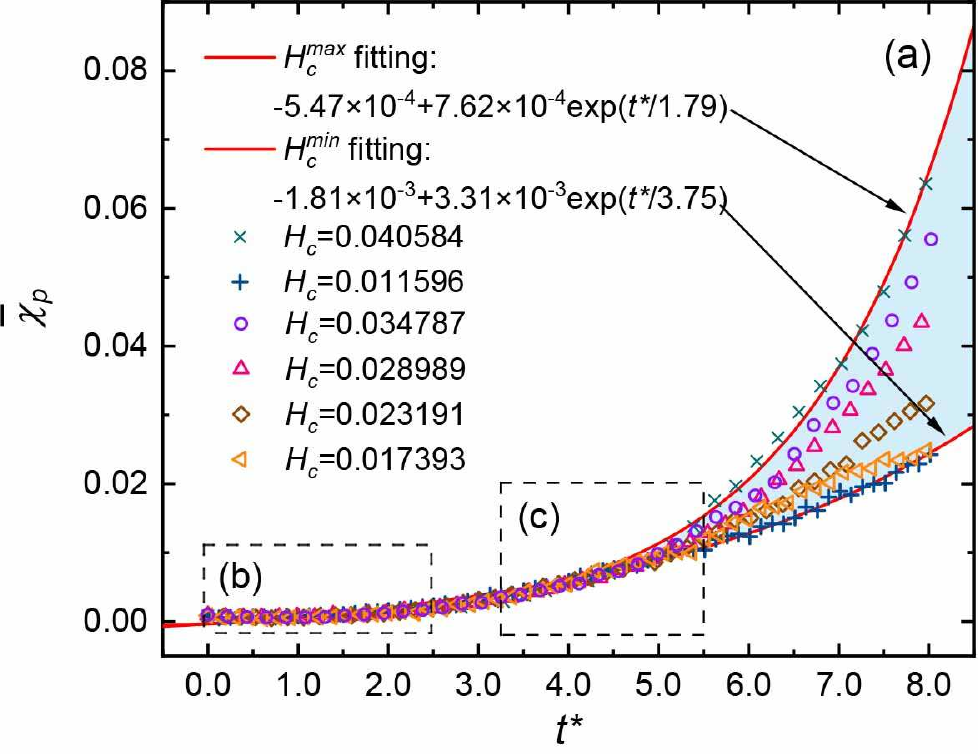}
    \end{minipage}
    \begin{minipage}{8cm}
        \includegraphics[width=8cm]{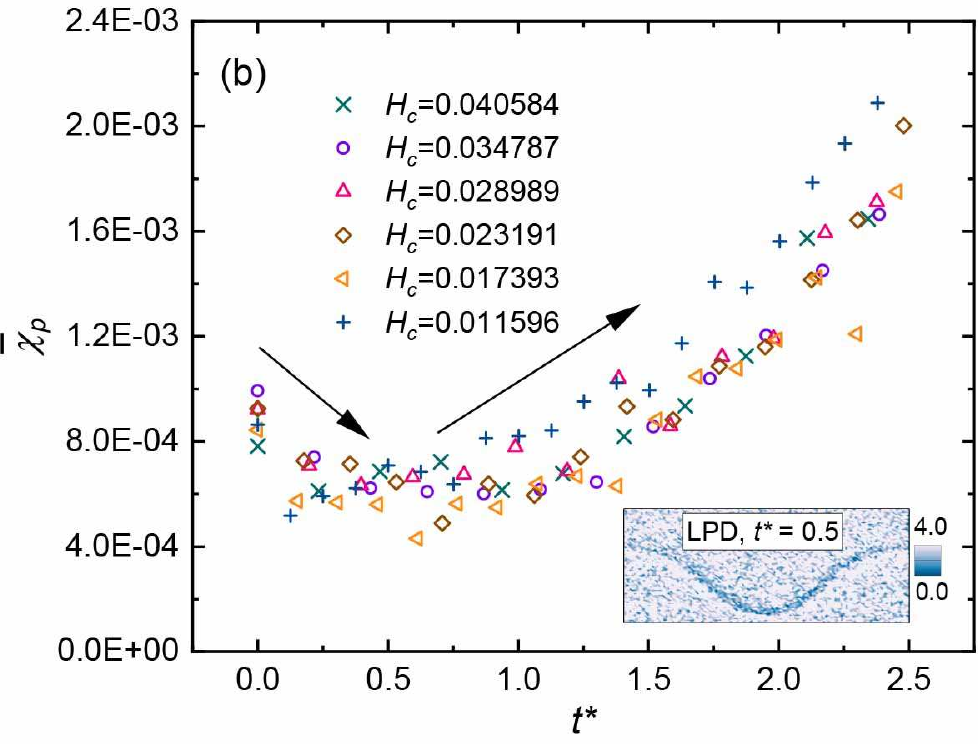}
    \end{minipage}
    \begin{minipage}{8cm}
        \includegraphics[width=8cm]{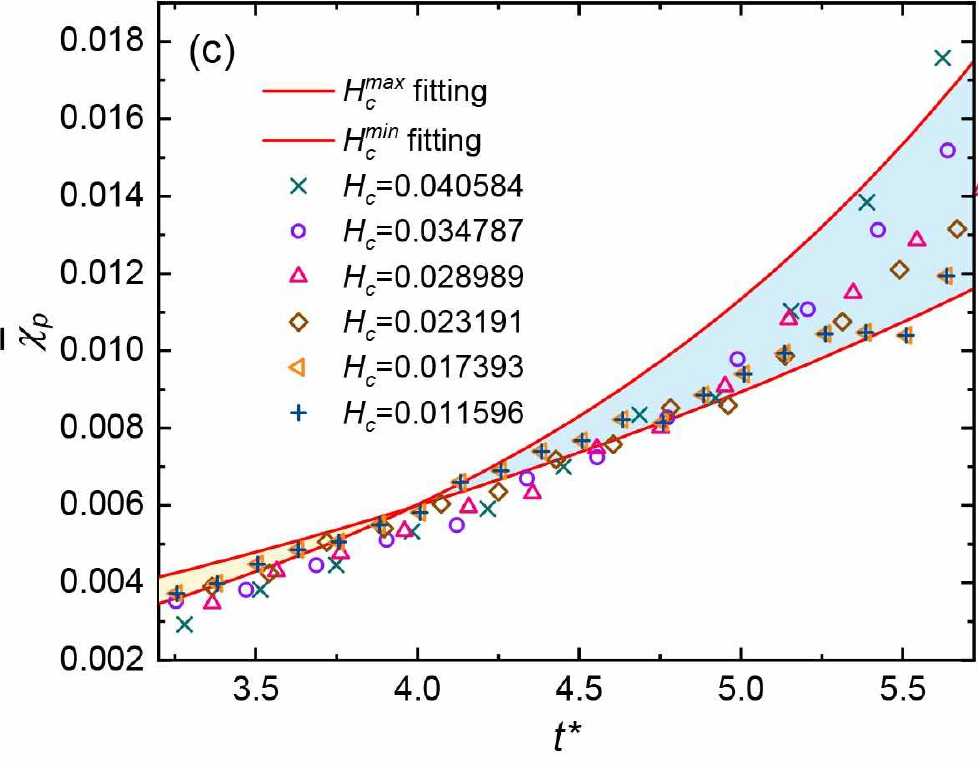}
    \end{minipage}
    \caption{Profile of average mixedness versus time under different compression factors. (a) an overall profile; (b) initial stage of RTI; the inner figure shows tracer distribution at $t^{*} = 0.5$ with a sparse band in the middle; (c) $t^{*}$ ranges from 3.5 to 5.5, during which the mixedness curves of high and low compression factor intersect.}
    \label{fig:cpsmix}
\end{figure*}

Firstly, the effect of compressibility on mixing is shown by Figure \ref{fig:cpsmix}, which provides the average mixedness at different times. The mixedness increases with time under each compression factor in Figure \ref{fig:cpsmix}(a). For a long-range from the initial time, curves of different compression factors show relatively close value, but with the increase of mixedness accelerates, curves of high compression show larger mixedness than low compression curves. Especially, the curve with the highest $H_c$ stands out at $t=8.0$. In this case, high compressibility may enhance the mixedness at the late stage of RTI mixing.

However, the mixedness at the late stage is too large compared with initial values. Two ranges of $t^{*}$ at the early time are selected to help us better understand the mixing process, as shown by Figure \ref{fig:cpsmix}(b) and (c). At the beginning process (see Figure \ref{fig:cpsmix}(b)), there is an obvious decline of mixedness before $t^{*} = 0.5$. According to previous analysis, the mixing starts from the interface, so the penetration of two fluids may be weakened at this stage. A tracer distribution pattern at $t^{*} = 0.5$ shows that the local tracer density is small around the interface, indicating each fluid is retreating to avoid further contact at the interface. After $t^{*} = 0.5$, due to the strong effect of acceleration, RTI continues to develop, but a decline-ascent pattern of mixedness is recorded.

\citet{Lai_2016} pointed that the effect of compressibility on the development of nonequilibrium in RTI is not monotonic destabilization \citep{Xue_pop_2010}, but with two stages, which initially inhibits and later enhances the development. On RTI mixing, the effect of compressibility is also with two stages, as shown by the intersection of $H_c^{max}$ and $H_c^{min}$ curves in Figure \ref{fig:cpsmix}(c). After the startup, the mixedness under high compressibility is lower than that of low compressibility, which is opposite to the relationship at the late stage. In this case, the curve of high compressibility catches up with the low compressibility curve at about $t^{*} = 3.9$.

\begin{figure*}
	\centerline{\includegraphics[width=14cm]{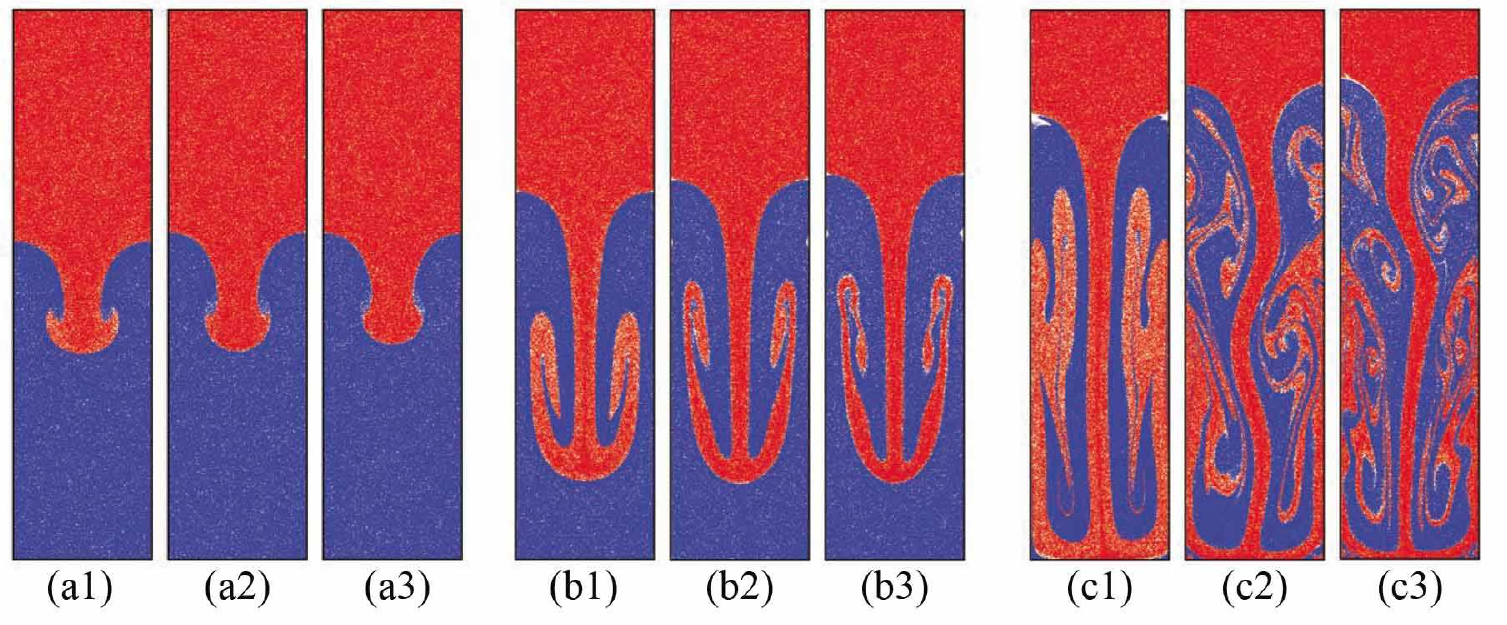}}
	\caption{Delineation of RTI flow and mixing by tracers under different compressibility; (1)-(3) indicate different compression factors, $H_c = 0.011596$, 0.028989, and 0.040584 respectively; (a)-(c) indicate different time, $t^{*} = 2.0$, 4.0, and 7.0 respectively.}
	\label{fig:cpspttn}
\end{figure*}

In order to understand the two stages of the effect of compressibility, Figure \ref{fig:cpspttn} demonstrates the tracer patterns of RTI flow, with three selected time points, $t^{*} = 2.0$, 4.0, and 7.0, approximately located at the first stage, the transition stage, and the late stage of mixedness development; three compression factors, $H_c = 0.011596$, 0.028989, and 0.040584, which are the minimum, a medium, and the maximum value in Figure \ref{fig:cpsmix}. At $t^{*} = 2.0$, see Figures \ref{fig:cpspttn}(a1) to (a3), the low compressibility system shows spike growing faster as well as larger deformation. This difference in large structure (spike, bubble, and etc.) leads to the greater mixedness of high compressibility system. At the transition point, $t^{*} = 4.0$, see Figures \ref{fig:cpspttn}(b1) to (b3), although fluids penetrate each other, the flow pattern in low compressibility system is simpler, without highly expanded small structures. However, in high compressibility system, the spike is elongated and unstably deforms, leading to complex flow patterns. At the late stage ($t^{*} = 7.0$), see Figures \ref{fig:cpspttn}(c1) to (c3), distinct difference is found between three patterns from left to right, corresponding to low to high compressibility. Compared with a low compressibility system, the flow in a high compressibility system is turbulent, associated with many small drops and tiny vortices. Contrarily, there is no evident dissociation of fluids in a low compressibility system. In the late stage, turbulence enhances the mixing while a more disordered state corresponds to the greater mixedness.

\begin{figure}
	\centerline{\includegraphics[width=8cm]{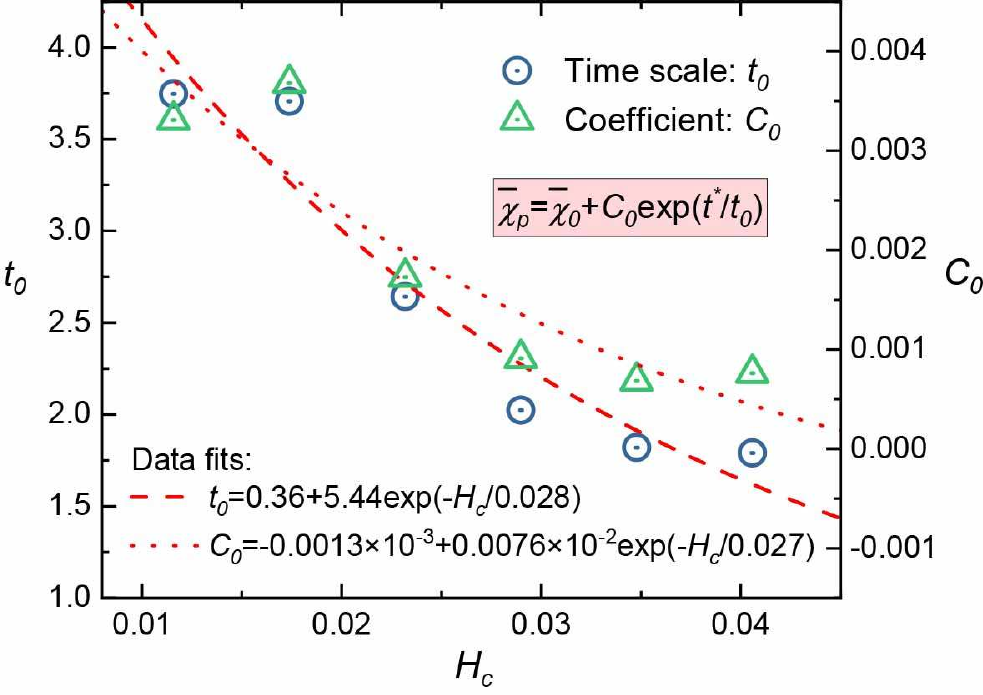}}
	\caption{Characteristic time scale $t_0$ and characteristic mixedness $C_0$ change with compression factor.}
	\label{fig:cpsscl}
\end{figure}

Fitting different curves in Figure \ref{fig:cpsmix} with an exponential function, $\overline{\chi}_{p} = \overline{\chi}_{0} + C_0 \cdot \exp \left( t^{*}/t_0 \right) $, where $\overline{\chi}_{0}$, $C_0$, and $t_0$ are fitting parameters. Mode of mixing in different compressibility systems is analyzed. In the fitting function, $t_0$ is the characteristic time scale of the mixing, and $C_0$ is a characteristic mixedness. Both parameters decline with the increase of compression factor $H_c$, as shown by Figure \ref{fig:cpsscl}. Higher compressibility would lead to a twofold effect, a higher mixing rate ($1/t_0$) and a lower characteristic mixedness ($C_0$), the two together govern the mixedness. In a short period of time after the start of RTI, mixedness is mainly controlled by $C_0$; thus, compressibility inhibits the mixing. As time increases, the exponential term ($\exp({t^{*}/t_0})$) dominates the changes of mixedness, so compressibility enhances the mixing at the late stage.

Intuitively, fluids with low compressibility are ``hard'' and those with high compressibility are ``soft''. At the early stage, a ``hard'' one has relatively high penetrating ability and leads to the quick formation of large structures, which contributes to the mixedness. At the late stage, small structures tend to generate in the ``soft'' system, which then leads to high mixedness.

\subsection{Viscous effect on RTI mixing}\label{subsec:visco}
Due to in many cases RTI exists in an extreme physical condition with drastic changes \citep{Weber_pre_2014}, the viscosity of fluid is not constant and would furtherly impact RTI development. Generally, viscosity is considered a stabilization factor on RTI at the start-up stage as well as the reacceleration stage, due to viscous force acts as resistance and inhibits the development of KHI in late-stage which then inhibits the RTI \citep{Sohn_2009, Chen_2018}. However, the role of viscosity is not simple because, after the reacceleration stage, viscosity could lead to complex flow patterns which contribute to the mixing of fluids \citep{Hu_pof_2019}. Therefore, we intend to discuss the role viscosity plays in mixing.

Ahead of the discussion on viscous effect, it is noted that in the single-relaxation-time discrete Boltzmann model (SRT-DBM) with the ideal gas equation, the product of relaxation time $\tau$ and temperature $T$ governs the viscosity,
\begin{equation}
    \mu=\tau \rho RT=\tau P
    \label{eq:viscosity}
\end{equation}

Due to the pressure is changeable during the RTI development, the initial pressure at the interface, $p_0 = 1.0$, is utilized to determine the viscosity. Different values of relaxation time $\tau$ are employed to alter the viscosity. Additionally, as an intrinsic physical parameter of fluid, the relaxation time $\tau$ links with Knudsen number $H_{\tau}$ either, which indicates the nonequilibrium state of the fluid and has already been shown in equation \eqref{eq:Htau}.

\begin{figure*}
	\centering
	\begin{minipage}{8cm}
		\includegraphics[width=8cm]{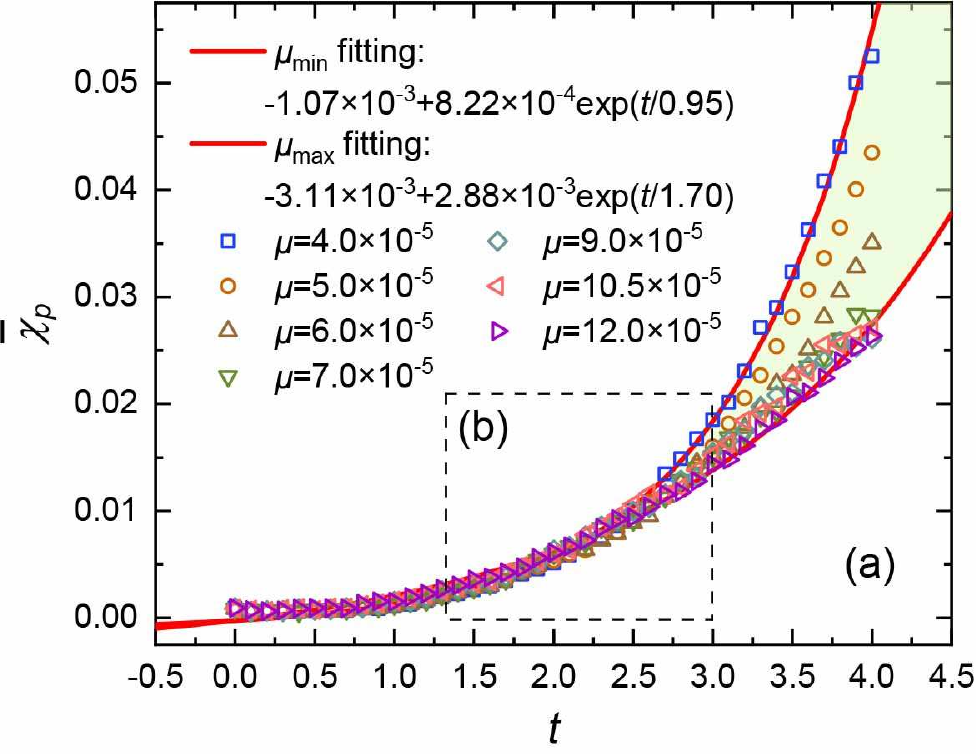}
	\end{minipage}
	\begin{minipage}{8cm}
		\includegraphics[width=8cm]{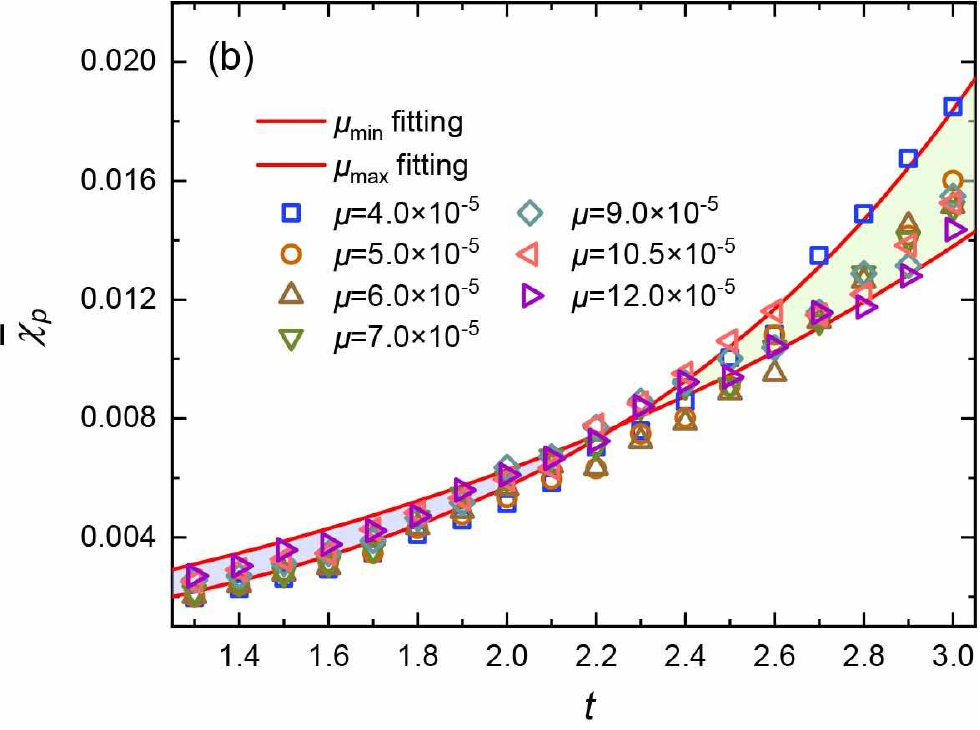}
	\end{minipage}
	\begin{minipage}{8cm}
		\includegraphics[width=8cm]{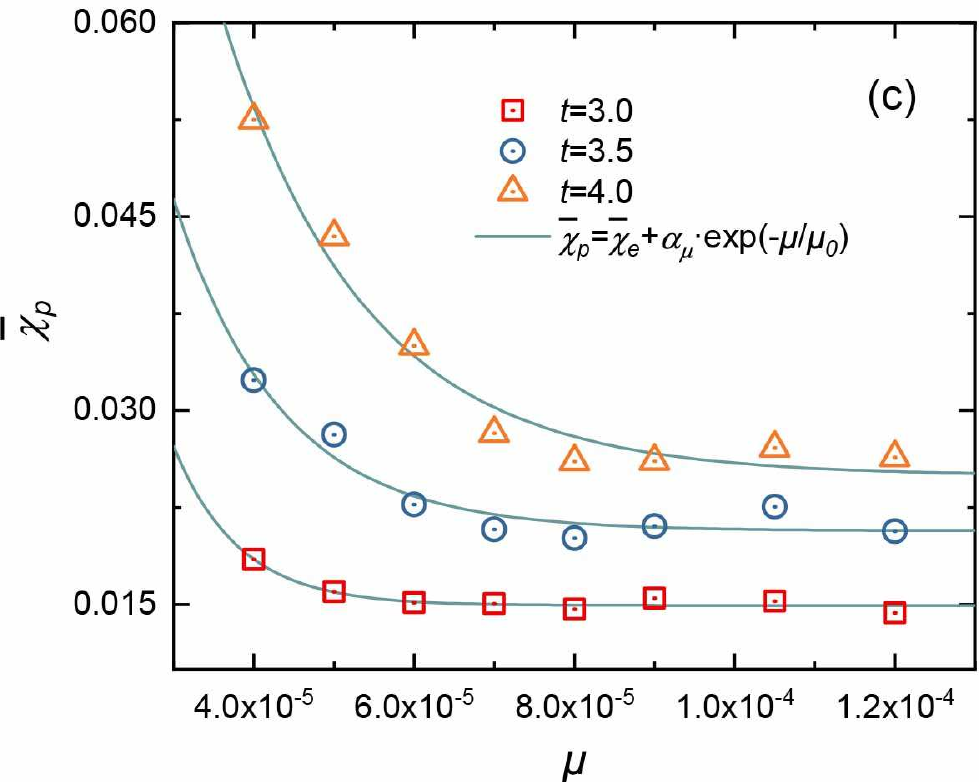}
	\end{minipage}
	\caption{Profile of average mixedness versus time under different viscosity. (a) an overall profile; (b) $t$ ranges from 1.3 to 3.0, during which the mixedness curves of high and low viscosity intersect; (c) the relationship between mixedness and viscosity at specific times.}
	\label{fig:vismix}
\end{figure*}

Figure \ref{fig:vismix} shows the average mixedness obtained from RTI Mixing under different viscosity, with $\mu$ ranges from $4.0 \times 10^{-5}$ to $12.0 \times 10^{-5}$. In Figure \ref{fig:vismix}(a), all curves are close to each other at the beginning, but at late time, curves of low viscosity show greater mixedness than that of high viscosity. The late-stage result is in good agreement with the common understanding that viscosity acts as an inhibition effect on RTI development, which could be extended to RTI induced mixing here. When magnifying curves of average mixedness near the start of RTI flow, a similar decline-ascent relationship is found between mixedness and time, which is in accord with Figure \ref{fig:cpsmix}(b). After this period, an intersection of different curves is found in Figure \ref{fig:vismix}(b), indicating viscosity is also not a monotonic effect on mixing. From the result, at the initial stage, high viscosity enhances the mixing, but in the late stage, high viscosity inhibits the mixing. An interesting phenomenon at the late stage is that, for a specific time, the mixedness does not continue to decrease with the increase of viscosity, as shown by Figure \ref{fig:vismix}(c). The relationship between mixedness and viscosity at three time points, $t = 3.0$, 3.5, and 4.0, are demonstrated in this figure. The mixedness slowly drops and tends to a constant value with an increment of viscosity. The saturation of mixedness with viscosity indicates that large viscosity can suppress the RTI mixing and the effect is diminishing with the further increase of viscosity. However, at different times, the saturation mixedness is different and mixedness still increases with time.

\begin{figure*}
	\centerline{\includegraphics[width=14cm]{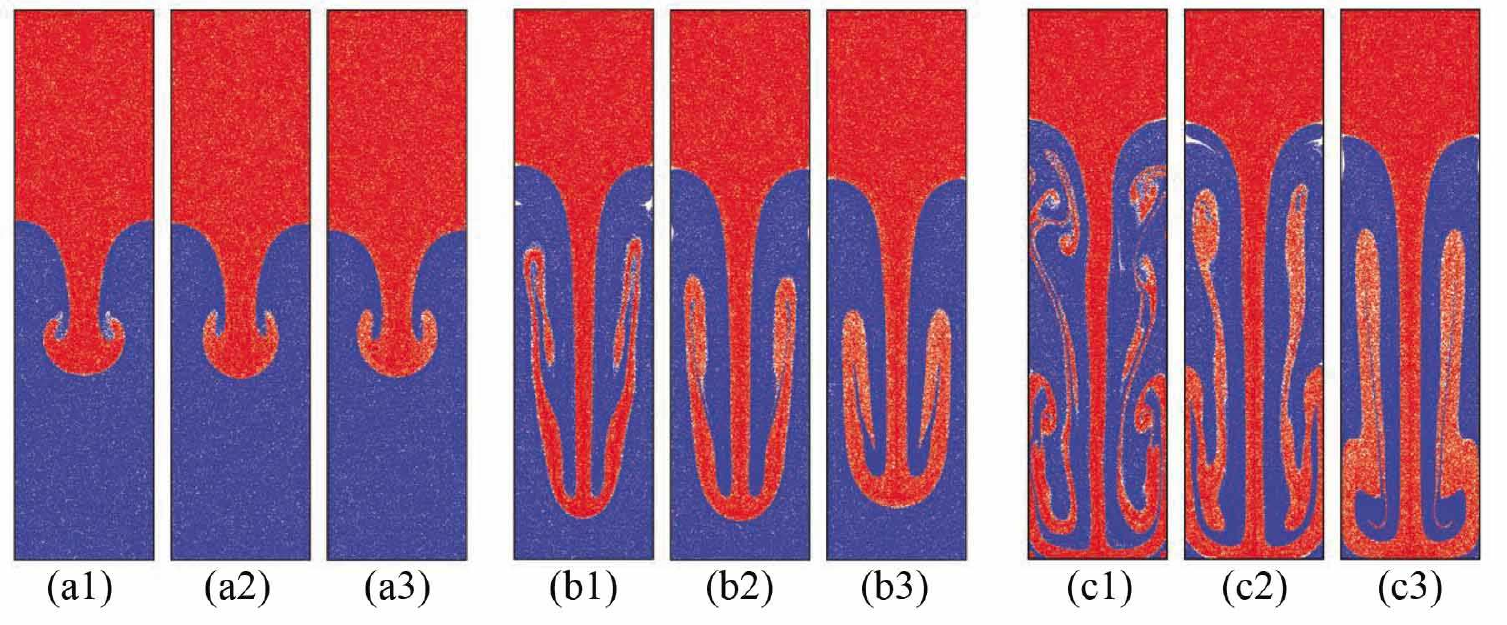}}
	\caption{Delineation of RTI flow and mixing by tracers under different viscosity; (1)-(3) indicate different viscosity, $\mu = 4 \times 10^{-5}$, $7 \times 10^{-5}$, and $12 \times 10^{-5}$ respectively; (a)-(c) indicate different time, $t = 1.2$, 2.2, and 3.0 respectively.}
	\label{fig:vispttn}
\end{figure*}

Figure \ref{fig:vispttn} is made for an illustration of viscous effect on mixing at different stages. Three time points, $t = 1.2$, 2.2, and 3.0 are selected, which approximately locate at the initial stage, the transition stage, and the late state respectively. Three different viscosity, $\mu = 4 \times 10^{-5}$, $7 \times 10^{-5}$, and $12 \times 10^{-5}$) are employes in the delineation. At $t = 1.2$, according to Figure \ref{fig:vispttn}(a), the spike in the high viscosity system is integral and with more contact with the bubble in the vortex region, which lead to higher value of mixedness. In this stage, we could also attribute the mixing to large-scale development. At $t = 2.2$, the mixedness of low viscosity system starts to become larger than that of high viscosity system. We can observe the elongated structure of fluids in Figure \ref{fig:vispttn}(b1), but in Figure \ref{fig:vispttn}(b3), which shows high viscous fluids, the development is constrained. Significant differences are observed at $t = 3.0$ through Figure \ref{fig:vispttn}(c), when the mixedness of low viscosity system fully transcends that of high viscosity system. At the time, as more tiny flow structures break down from mainstream and drops dissociate from the main body in the low viscosity system, the mixedness of low viscosity system significantly increases. In contrast, in high viscosity system, large structures are maintained. The results of RTI mixing at the late stage are in accord with the conclusion made by \citet{Liang_2019} that relatively high Reynolds number (corresponding to low viscosity in this paper) leads to high growth rate of RTI and generation of dissociated tiny fluid structures. However, detailed results presented above can make compensation for this understanding, that initially, viscosity is a favorable effect for mixing.

\begin{figure}
	\centerline{\includegraphics[width=8cm]{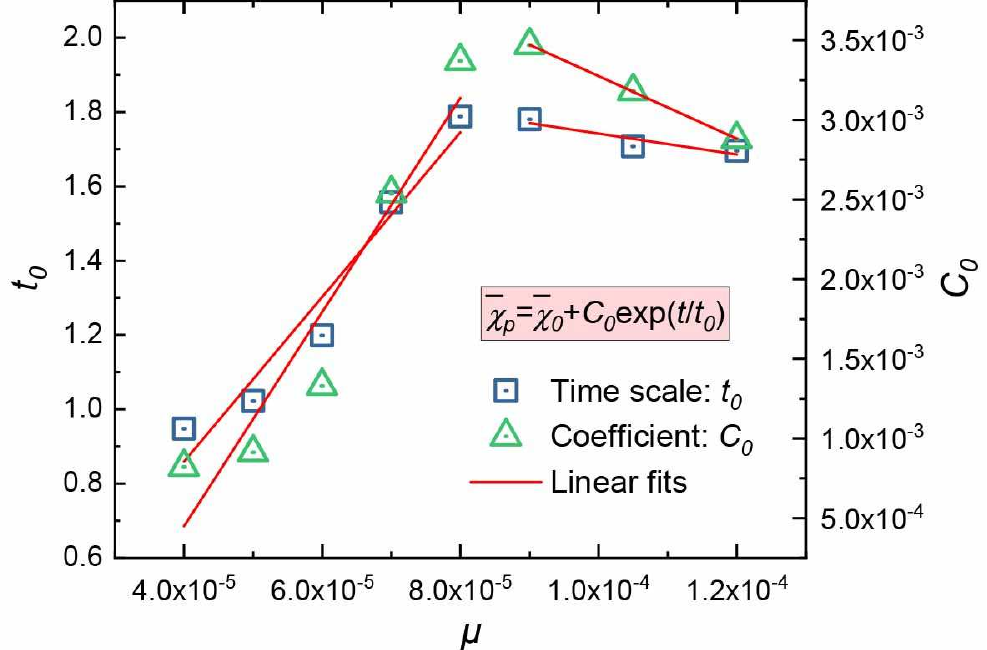}}
	\caption{Characteristic time scale $t_0$ and characteristic mixedness $C_0$ changes with viscosity.}
	\label{fig:visscl}
\end{figure}

A scale analysis based on optimal exponential fitting ($\overline{\chi}_{p} = \overline{\chi}_{0} + C_0 \cdot \exp \left( t^{*}/t_0 \right) $) of average mixedness over viscosity is conducted, and Figure \ref{fig:visscl} shows the characteristic time scale ($t_0$) and characteristic mixedness ($C_0$). The curves of two parameters both can be separated into two parts, which is different from the compressibility. The time scale increases with viscosity initially and then becomes nearly constant with slight decrease. Characteristic mixedness shows similar relationship with viscosity. At late stage, the stabilization of two parameters leads to the saturation effect of mixedness. From the scale analysis, the effect of viscosity is not only attributed to the transition of dominant role between the time scale and characteristic mixedness, but also to the different stages of $t_0$ and $C_0$ respectively.

The two-stage effect of viscosity is also caused by the separated development of large structure and small structure at two stages. A fluid system with high viscosity is considered ``hard'' and one with low viscosity is considered ``soft''. At the first stage, a ``hard'' one has a relatively higher penetrating ability and leads to higher mixedness, while a ``soft'' one has low penetrating ability as well as low mixedness. However, at the late stage, more tiny structures tend to generate in the ``soft'' system due to it is more delicate than the ``hard'' one, which leads to higher mixedness. Additionally, a ``hard'' enough system can successfully suppress the generation of tiny structures, which causes a saturation effect of viscosity on mixedness.

\section{Conclusions}\label{sec:conclusions}
In this paper, tracers are introduced into the DBM simulation of compressible RTI flow. The tracers enable observation of fine structures with clear interfaces in the late RTI mixing stage. In the velocity space of tracers, interesting distribution patterns are shown at different stages. These patterns bring quite dense information for the RTI research, which opens a new way for analyzing and accessing significantly deep insights into the flow system. With the aid of tracing the interface as well as interfacial physical quantities, various nonequilibrium behaviors on the interface of fluids are systematically investigated.

A mixedness $\chi_p$ is defined by the local spatial distribution of tracers. With statistics, the distribution of vertically averaged mixedness $\widetilde{\chi}_p$ along the horizontal direction shows an interesting perspective to study the occurrence of KHI as well as evaluation of KHI intensity.

The effect of two important physical factors, compressibility and viscosity, are explored for RTI mixing. Both of them show two-stage effects. Specifically, high compressibility and low viscosity initially inhibits and later enhances the RTI mixing. The effect can be roughly understood via the following physical picture. In the initial period, the mixing is mainly induced by the evolution of large structures, while at the late period the mixing is mainly induced by the generation of small structures. Low compressibility and/or high viscosity make the system ``hard'' which favors the initial evolution of large structures. High compressibility and/or low viscosity make the system ``soft'' which favors the generation of small structures in the late stage.

The field averaged mixedness $\overline{\chi}_p$ increases nearly exponentially with time, which indicates the existence of a characteristic time scale $t_0$. The time scale $t_0$ shows different dependences on compressibility and viscosity. It monotonically decreases with compressibility, while shows a two-stage dependence on viscosity. It firstly increases and then slightly decreases with viscosity, which means there is an appropriate viscosity for the system to get minimum mixedness. It is indicated that one can inhibit the mixedness via adjusting system viscosity. At the late stage, for a fixed time, the field averaged mixedness shows an exponential decrease with the viscosity, which indicates two points, (i) there exists a characteristic viscosity $\mu_0$ and (ii) the mixedness will not show a meaningful decrease with viscosity after a certain value reaches.

\section*{Data availability}
The data that support the findings of this study are available from the corresponding author upon reasonable request.

\begin{acknowledgments}
Thanks to Dr. Yudong Zhang, Dr. Kaige Zhao, Prof. Lihao Zhao, Mr. Yiming Shan, and Mr. Jiahui Song for their kind help. Aiguo acknowledges the support from the National Natural Science Foundation of China (Grant Nos. 11772064 and 11574390), CAEP Foundation (Grant No. CX2019033), Science Challenge Project (Grant No. JCKY2016212A501), and the opening project of State Key Laboratory of Explosion Science and Technology (Beijing Institute of Technology, Grant No. KFJJ21-16M). Huilin acknowledges the support from Natural Science Foundation of Fujian Provinces (Grant No. 2018J01654).
\end{acknowledgments}

\appendix
\section{Mixedness based on different statistical cells}\label{appA}

\begin{figure}
    \centering
    \begin{minipage}{8cm}
        \includegraphics[width=8cm]{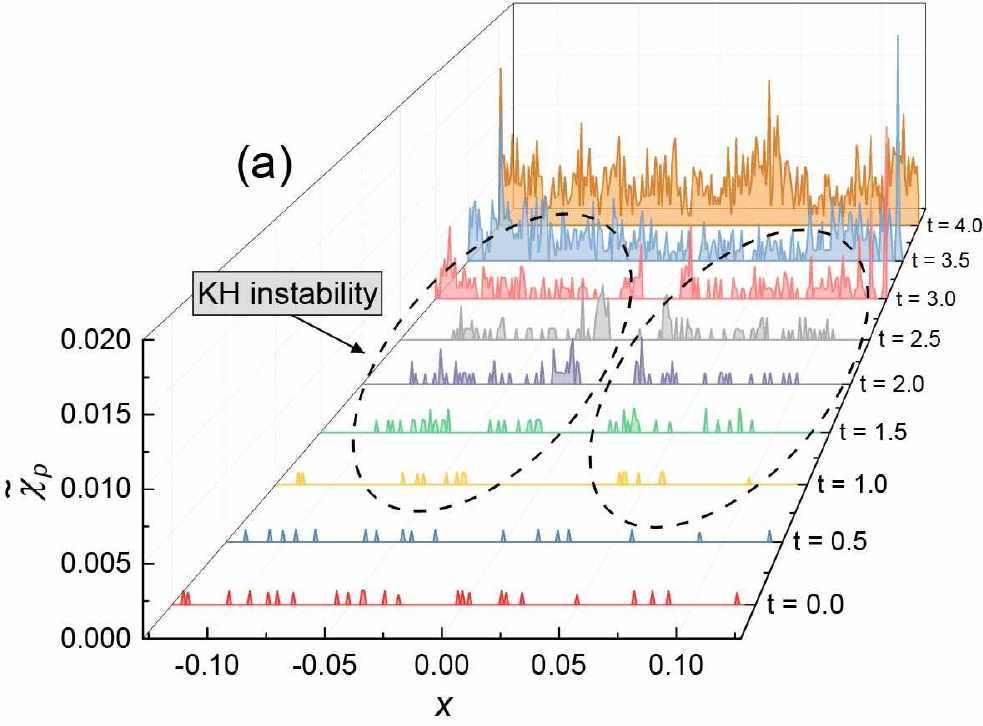}
    \end{minipage}
    \begin{minipage}{8cm}
        \includegraphics[width=8cm]{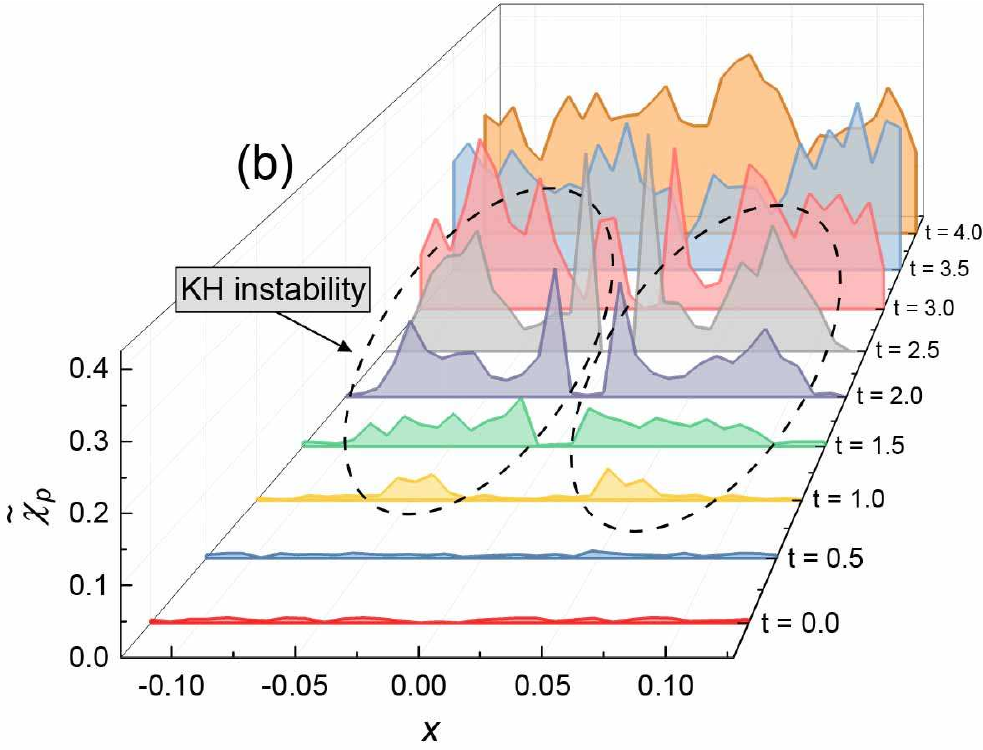}
    \end{minipage}

    \caption{Profiles of average mixedness $\widetilde{\chi}_p$ along the horizontal direction at different time. Circles emphasize mixing regions with KHI. Results are obtained based on (a) $n = 1.0$ and (b) $n = 8.0$ statistical cells.}
    \label{fig:a1}
\end{figure}

Profile of mixedness which is averaged along the vertical direction is obtained based on two statistical cells, $n=1.0$ and 8.0 respectively, shown by Figures \ref{fig:a1} (a) and (b). It is found that the general patterns are in accord with that in Figure \ref{fig:vm}. However, with a small statistical cell, the distribution curve of mixedness fluctuates, while with a large statistical cell, the distribution curve is over-smoothed. An intermediate-sized statistical cell can take advantages and avoid disadvantages of both, better delineating the mixing process as shown in the main text.

\section*{References}
\bibliography{pof-reference}

\begin{thebibliography}{74}%
\makeatletter
\providecommand \@ifxundefined [1]{%
 \@ifx{#1\undefined}
}%
\providecommand \@ifnum [1]{%
 \ifnum #1\expandafter \@firstoftwo
 \else \expandafter \@secondoftwo
 \fi
}%
\providecommand \@ifx [1]{%
 \ifx #1\expandafter \@firstoftwo
 \else \expandafter \@secondoftwo
 \fi
}%
\providecommand \natexlab [1]{#1}%
\providecommand \enquote  [1]{``#1''}%
\providecommand \bibnamefont  [1]{#1}%
\providecommand \bibfnamefont [1]{#1}%
\providecommand \citenamefont [1]{#1}%
\providecommand \href@noop [0]{\@secondoftwo}%
\providecommand \href [0]{\begingroup \@sanitize@url \@href}%
\providecommand \@href[1]{\@@startlink{#1}\@@href}%
\providecommand \@@href[1]{\endgroup#1\@@endlink}%
\providecommand \@sanitize@url [0]{\catcode `\\12\catcode `\$12\catcode
  `\&12\catcode `\#12\catcode `\^12\catcode `\_12\catcode `\%12\relax}%
\providecommand \@@startlink[1]{}%
\providecommand \@@endlink[0]{}%
\providecommand \url  [0]{\begingroup\@sanitize@url \@url }%
\providecommand \@url [1]{\endgroup\@href {#1}{\urlprefix }}%
\providecommand \urlprefix  [0]{URL }%
\providecommand \Eprint [0]{\href }%
\providecommand \doibase [0]{http://dx.doi.org/}%
\providecommand \selectlanguage [0]{\@gobble}%
\providecommand \bibinfo  [0]{\@secondoftwo}%
\providecommand \bibfield  [0]{\@secondoftwo}%
\providecommand \translation [1]{[#1]}%
\providecommand \BibitemOpen [0]{}%
\providecommand \bibitemStop [0]{}%
\providecommand \bibitemNoStop [0]{.\EOS\space}%
\providecommand \EOS [0]{\spacefactor3000\relax}%
\providecommand \BibitemShut  [1]{\csname bibitem#1\endcsname}%
\let\auto@bib@innerbib\@empty
\bibitem [{\citenamefont {Rayleigh}(1883)}]{Rayleigh_1883}%
  \BibitemOpen
  \bibfield  {author} {\bibinfo {author} {\bibnamefont {Rayleigh}},\ }\bibfield
   {title} {\enquote {\bibinfo {title} {Investigation of the character of the
  equilibrium of an incompressible heavy fluid of variable density},}\ }\href
  {\doibase {}} {\bibfield  {journal} {\bibinfo  {journal} {Proceedings of the
  London Mathematical Society}\ }\textbf {\bibinfo {volume} {14}},\ \bibinfo
  {pages} {170--177} (\bibinfo {year} {1883})}\BibitemShut {NoStop}%
\bibitem [{\citenamefont {Taylor}(1950)}]{Taylor_1950}%
  \BibitemOpen
  \bibfield  {author} {\bibinfo {author} {\bibfnamefont {G.~I.}\ \bibnamefont
  {Taylor}},\ }\bibfield  {title} {\enquote {\bibinfo {title} {The instability
  of liquid surfaces when accelerated in a direction perpendicular to their
  planes. {I}},}\ }\href {\doibase {}} {\bibfield  {journal} {\bibinfo
  {journal} {Proceedings of the Royal Society of London. Series A. Mathematical
  and Physical Sciences}\ }\textbf {\bibinfo {volume} {201}},\ \bibinfo {pages}
  {192--196} (\bibinfo {year} {1950})}\BibitemShut {NoStop}%
\bibitem [{\citenamefont {Lewis}(1950)}]{Lewis_1950}%
  \BibitemOpen
  \bibfield  {author} {\bibinfo {author} {\bibfnamefont {D.~J.}\ \bibnamefont
  {Lewis}},\ }\bibfield  {title} {\enquote {\bibinfo {title} {The instability
  of liquid surfaces when accelerated in a direction perpendicular to their
  planes. {II}},}\ }\href {\doibase {}} {\bibfield  {journal} {\bibinfo
  {journal} {Proceedings of the Royal Society of London. Series A. Mathematical
  and Physical Sciences}\ }\textbf {\bibinfo {volume} {202}},\ \bibinfo {pages}
  {81--96} (\bibinfo {year} {1950})}\BibitemShut {NoStop}%
\bibitem [{\citenamefont {Ribeyre}, \citenamefont {Tikhonchuk},\ and\
  \citenamefont {Bouquet}(2004)}]{Ribeyre_2004}%
  \BibitemOpen
  \bibfield  {author} {\bibinfo {author} {\bibfnamefont {X.}~\bibnamefont
  {Ribeyre}}, \bibinfo {author} {\bibfnamefont {V.}~\bibnamefont {Tikhonchuk}},
  \ and\ \bibinfo {author} {\bibfnamefont {S.}~\bibnamefont {Bouquet}},\
  }\bibfield  {title} {\enquote {\bibinfo {title} {Compressible
  {Rayleigh--Taylor} instabilities in supernova remnants},}\ }\href {\doibase
  {}} {\bibfield  {journal} {\bibinfo  {journal} {Physics of Fluids}\ }\textbf
  {\bibinfo {volume} {16}},\ \bibinfo {pages} {4661--4670} (\bibinfo {year}
  {2004})}\BibitemShut {NoStop}%
\bibitem [{\citenamefont {Lev}\ and\ \citenamefont {Hager}(2008)}]{Lev_2008}%
  \BibitemOpen
  \bibfield  {author} {\bibinfo {author} {\bibfnamefont {E.}~\bibnamefont
  {Lev}}\ and\ \bibinfo {author} {\bibfnamefont {B.~H.}\ \bibnamefont
  {Hager}},\ }\bibfield  {title} {\enquote {\bibinfo {title}
  {{Rayleigh--Taylor} instabilities with anisotropic lithospheric viscosity},}\
  }\href {\doibase {}} {\bibfield  {journal} {\bibinfo  {journal} {Geophysical
  Journal International}\ }\textbf {\bibinfo {volume} {173}},\ \bibinfo {pages}
  {806--814} (\bibinfo {year} {2008})}\BibitemShut {NoStop}%
\bibitem [{\citenamefont {Vinningland}\ \emph {et~al.}(2007)\citenamefont
  {Vinningland}, \citenamefont {Johnsen}, \citenamefont {Flekk{\o}y},
  \citenamefont {Toussaint},\ and\ \citenamefont
  {M{\aa}l{\o}y}}]{Vinningland_2007}%
  \BibitemOpen
  \bibfield  {author} {\bibinfo {author} {\bibfnamefont {J.~L.}\ \bibnamefont
  {Vinningland}}, \bibinfo {author} {\bibfnamefont {{\O}.}~\bibnamefont
  {Johnsen}}, \bibinfo {author} {\bibfnamefont {E.~G.}\ \bibnamefont
  {Flekk{\o}y}}, \bibinfo {author} {\bibfnamefont {R.}~\bibnamefont
  {Toussaint}}, \ and\ \bibinfo {author} {\bibfnamefont {K.~J.}\ \bibnamefont
  {M{\aa}l{\o}y}},\ }\bibfield  {title} {\enquote {\bibinfo {title} {Granular
  {Rayleigh-Taylor} instability: {Experiments} and simulations},}\ }\href
  {\doibase {}} {\bibfield  {journal} {\bibinfo  {journal} {Physical review
  letters}\ }\textbf {\bibinfo {volume} {99}},\ \bibinfo {pages} {048001}
  (\bibinfo {year} {2007})}\BibitemShut {NoStop}%
\bibitem [{\citenamefont {Lindl}\ \emph {et~al.}(2004)\citenamefont {Lindl},
  \citenamefont {Amendt}, \citenamefont {Berger}, \citenamefont {Glendinning},
  \citenamefont {Glenzer}, \citenamefont {Haan}, \citenamefont {Kauffman},
  \citenamefont {Landen},\ and\ \citenamefont {Suter}}]{Lindl_2004}%
  \BibitemOpen
  \bibfield  {author} {\bibinfo {author} {\bibfnamefont {J.~D.}\ \bibnamefont
  {Lindl}}, \bibinfo {author} {\bibfnamefont {P.}~\bibnamefont {Amendt}},
  \bibinfo {author} {\bibfnamefont {R.~L.}\ \bibnamefont {Berger}}, \bibinfo
  {author} {\bibfnamefont {S.~G.}\ \bibnamefont {Glendinning}}, \bibinfo
  {author} {\bibfnamefont {S.~H.}\ \bibnamefont {Glenzer}}, \bibinfo {author}
  {\bibfnamefont {S.~W.}\ \bibnamefont {Haan}}, \bibinfo {author}
  {\bibfnamefont {R.~L.}\ \bibnamefont {Kauffman}}, \bibinfo {author}
  {\bibfnamefont {O.~L.}\ \bibnamefont {Landen}}, \ and\ \bibinfo {author}
  {\bibfnamefont {L.~J.}\ \bibnamefont {Suter}},\ }\bibfield  {title} {\enquote
  {\bibinfo {title} {The physics basis for ignition using indirect-drive
  targets on the {National Ignition Facility}},}\ }\href {\doibase {}}
  {\bibfield  {journal} {\bibinfo  {journal} {Physics of Plasmas}\ }\textbf
  {\bibinfo {volume} {11}},\ \bibinfo {pages} {339--491} (\bibinfo {year}
  {2004})}\BibitemShut {NoStop}%
\bibitem [{\citenamefont {Edwards}\ \emph {et~al.}(2013)\citenamefont
  {Edwards}, \citenamefont {Patel}, \citenamefont {Lindl}, \citenamefont
  {Atherton}, \citenamefont {Glenzer}, \citenamefont {Haan}, \citenamefont
  {Kilkenny}, \citenamefont {Landen}, \citenamefont {Moses},\ and\
  \citenamefont {Nikroo}}]{Edwards_2013}%
  \BibitemOpen
  \bibfield  {author} {\bibinfo {author} {\bibfnamefont {M.}~\bibnamefont
  {Edwards}}, \bibinfo {author} {\bibfnamefont {P.}~\bibnamefont {Patel}},
  \bibinfo {author} {\bibfnamefont {J.}~\bibnamefont {Lindl}}, \bibinfo
  {author} {\bibfnamefont {L.}~\bibnamefont {Atherton}}, \bibinfo {author}
  {\bibfnamefont {S.}~\bibnamefont {Glenzer}}, \bibinfo {author} {\bibfnamefont
  {S.}~\bibnamefont {Haan}}, \bibinfo {author} {\bibfnamefont {J.}~\bibnamefont
  {Kilkenny}}, \bibinfo {author} {\bibfnamefont {O.}~\bibnamefont {Landen}},
  \bibinfo {author} {\bibfnamefont {E.}~\bibnamefont {Moses}}, \ and\ \bibinfo
  {author} {\bibfnamefont {A.}~\bibnamefont {Nikroo}},\ }\bibfield  {title}
  {\enquote {\bibinfo {title} {Progress towards ignition on the {National
  Ignition Facility}},}\ }\href {\doibase {}} {\bibfield  {journal} {\bibinfo
  {journal} {Physics of Plasmas}\ }\textbf {\bibinfo {volume} {20}},\ \bibinfo
  {pages} {070501} (\bibinfo {year} {2013})}\BibitemShut {NoStop}%
\bibitem [{\citenamefont {Wang}\ \emph {et~al.}(2016)\citenamefont {Wang},
  \citenamefont {Ye}, \citenamefont {Wu}, \citenamefont {Liu}, \citenamefont
  {Zhang},\ and\ \citenamefont {He}}]{Wang_2016}%
  \BibitemOpen
  \bibfield  {author} {\bibinfo {author} {\bibfnamefont {L.}~\bibnamefont
  {Wang}}, \bibinfo {author} {\bibfnamefont {W.}~\bibnamefont {Ye}}, \bibinfo
  {author} {\bibfnamefont {J.}~\bibnamefont {Wu}}, \bibinfo {author}
  {\bibfnamefont {J.}~\bibnamefont {Liu}}, \bibinfo {author} {\bibfnamefont
  {W.}~\bibnamefont {Zhang}}, \ and\ \bibinfo {author} {\bibfnamefont
  {X.}~\bibnamefont {He}},\ }\bibfield  {title} {\enquote {\bibinfo {title} {A
  scheme for reducing deceleration-phase {Rayleigh--Taylor} growth in inertial
  confinement fusion implosions},}\ }\href {\doibase {}} {\bibfield  {journal}
  {\bibinfo  {journal} {Physics of Plasmas}\ }\textbf {\bibinfo {volume}
  {23}},\ \bibinfo {pages} {052713} (\bibinfo {year} {2016})}\BibitemShut
  {NoStop}%
\bibitem [{\citenamefont {Bradley}(2014)}]{Bradley_2014}%
  \BibitemOpen
  \bibfield  {author} {\bibinfo {author} {\bibfnamefont {P.}~\bibnamefont
  {Bradley}},\ }\bibfield  {title} {\enquote {\bibinfo {title} {The effect of
  mix on capsule yields as a function of shell thickness and gas fill},}\
  }\href {\doibase {}} {\bibfield  {journal} {\bibinfo  {journal} {Physics of
  Plasmas}\ }\textbf {\bibinfo {volume} {21}},\ \bibinfo {pages} {062703}
  (\bibinfo {year} {2014})}\BibitemShut {NoStop}%
\bibitem [{\citenamefont {Zhou}(2017{\natexlab{a}})}]{Zhou_PR1_2017}%
  \BibitemOpen
  \bibfield  {author} {\bibinfo {author} {\bibfnamefont {Y.}~\bibnamefont
  {Zhou}},\ }\bibfield  {title} {\enquote {\bibinfo {title} {{Rayleigh--Taylor}
  and {Richtmyer--Meshkov} instability induced flow, turbulence, and mixing.
  {I}},}\ }\href {\doibase https://doi.org/10.1016/j.physrep.2017.07.005}
  {\bibfield  {journal} {\bibinfo  {journal} {Physics Reports}\ }\textbf
  {\bibinfo {volume} {720-722}},\ \bibinfo {pages} {1--136} (\bibinfo {year}
  {2017}{\natexlab{a}})}\BibitemShut {NoStop}%
\bibitem [{\citenamefont {Zhou}(2017{\natexlab{b}})}]{Zhou_PR2_2017}%
  \BibitemOpen
  \bibfield  {author} {\bibinfo {author} {\bibfnamefont {Y.}~\bibnamefont
  {Zhou}},\ }\bibfield  {title} {\enquote {\bibinfo {title} {{Rayleigh--Taylor}
  and {Richtmyer--Meshkov} instability induced flow, turbulence, and mixing.
  {II}},}\ }\href {\doibase https://doi.org/10.1016/j.physrep.2017.07.008}
  {\bibfield  {journal} {\bibinfo  {journal} {Physics Reports}\ }\textbf
  {\bibinfo {volume} {723-725}},\ \bibinfo {pages} {1--160} (\bibinfo {year}
  {2017}{\natexlab{b}})}\BibitemShut {NoStop}%
\bibitem [{\citenamefont {Kull}(1991)}]{Kull_1991}%
  \BibitemOpen
  \bibfield  {author} {\bibinfo {author} {\bibfnamefont {H.~J.}\ \bibnamefont
  {Kull}},\ }\bibfield  {title} {\enquote {\bibinfo {title} {Theory of the
  {Rayleigh-Taylor} instability},}\ }\href {\doibase {}} {\bibfield  {journal}
  {\bibinfo  {journal} {Physics Reports}\ }\textbf {\bibinfo {volume} {206}},\
  \bibinfo {pages} {197--325} (\bibinfo {year} {1991})}\BibitemShut {NoStop}%
\bibitem [{\citenamefont {Goncharov}(2002)}]{Goncharov_2002}%
  \BibitemOpen
  \bibfield  {author} {\bibinfo {author} {\bibfnamefont {V.}~\bibnamefont
  {Goncharov}},\ }\bibfield  {title} {\enquote {\bibinfo {title} {Analytical
  model of nonlinear, single-mode, classical {Rayleigh--Taylor} instability at
  arbitrary {Atwood} numbers},}\ }\href {\doibase {}} {\bibfield  {journal}
  {\bibinfo  {journal} {Physical Review Letters}\ }\textbf {\bibinfo {volume}
  {88}},\ \bibinfo {pages} {134502} (\bibinfo {year} {2002})}\BibitemShut
  {NoStop}%
\bibitem [{\citenamefont {Mikaelian}(1998)}]{Mikaelian_1998}%
  \BibitemOpen
  \bibfield  {author} {\bibinfo {author} {\bibfnamefont {K.~O.}\ \bibnamefont
  {Mikaelian}},\ }\bibfield  {title} {\enquote {\bibinfo {title} {Analytic
  approach to nonlinear {Rayleigh--Taylor} and {Richtmyer--Meshkov}
  instabilities},}\ }\href {\doibase {}} {\bibfield  {journal} {\bibinfo
  {journal} {Physical Review Letters}\ }\textbf {\bibinfo {volume} {80}},\
  \bibinfo {pages} {508} (\bibinfo {year} {1998})}\BibitemShut {NoStop}%
\bibitem [{\citenamefont {Zhao}\ \emph {et~al.}(2018)\citenamefont {Zhao},
  \citenamefont {Wang}, \citenamefont {Xue}, \citenamefont {Ye}, \citenamefont
  {Wu}, \citenamefont {Ding},\ and\ \citenamefont {Zhang}}]{Zhao_2018}%
  \BibitemOpen
  \bibfield  {author} {\bibinfo {author} {\bibfnamefont {K.}~\bibnamefont
  {Zhao}}, \bibinfo {author} {\bibfnamefont {L.}~\bibnamefont {Wang}}, \bibinfo
  {author} {\bibfnamefont {C.}~\bibnamefont {Xue}}, \bibinfo {author}
  {\bibfnamefont {W.}~\bibnamefont {Ye}}, \bibinfo {author} {\bibfnamefont
  {J.}~\bibnamefont {Wu}}, \bibinfo {author} {\bibfnamefont {Y.}~\bibnamefont
  {Ding}}, \ and\ \bibinfo {author} {\bibfnamefont {W.}~\bibnamefont {Zhang}},\
  }\bibfield  {title} {\enquote {\bibinfo {title} {Thin layer model for
  nonlinear evolution of the {Rayleigh--Taylor} instability},}\ }\href
  {\doibase {}} {\bibfield  {journal} {\bibinfo  {journal} {Physics of
  Plasmas}\ }\textbf {\bibinfo {volume} {25}},\ \bibinfo {pages} {032708}
  (\bibinfo {year} {2018})}\BibitemShut {NoStop}%
\bibitem [{\citenamefont {Clark}\ and\ \citenamefont
  {Zhou}(2003)}]{Clark_2003}%
  \BibitemOpen
  \bibfield  {author} {\bibinfo {author} {\bibfnamefont {T.~T.}\ \bibnamefont
  {Clark}}\ and\ \bibinfo {author} {\bibfnamefont {Y.}~\bibnamefont {Zhou}},\
  }\bibfield  {title} {\enquote {\bibinfo {title} {Self-similarity of two flows
  induced by instabilities},}\ }\href {\doibase {}} {\bibfield  {journal}
  {\bibinfo  {journal} {Physical Review E}\ }\textbf {\bibinfo {volume} {68}},\
  \bibinfo {pages} {066305} (\bibinfo {year} {2003})}\BibitemShut {NoStop}%
\bibitem [{\citenamefont {Zhou}\ \emph {et~al.}(2019)\citenamefont {Zhou},
  \citenamefont {Clark}, \citenamefont {Clark}, \citenamefont
  {Gail~Glendinning}, \citenamefont {Aaron~Skinner}, \citenamefont
  {Huntington}, \citenamefont {Hurricane}, \citenamefont {Dimits},\ and\
  \citenamefont {Remington}}]{Zhou_pof_2019}%
  \BibitemOpen
  \bibfield  {author} {\bibinfo {author} {\bibfnamefont {Y.}~\bibnamefont
  {Zhou}}, \bibinfo {author} {\bibfnamefont {T.~T.}\ \bibnamefont {Clark}},
  \bibinfo {author} {\bibfnamefont {D.~S.}\ \bibnamefont {Clark}}, \bibinfo
  {author} {\bibfnamefont {S.}~\bibnamefont {Gail~Glendinning}}, \bibinfo
  {author} {\bibfnamefont {M.}~\bibnamefont {Aaron~Skinner}}, \bibinfo {author}
  {\bibfnamefont {C.~M.}\ \bibnamefont {Huntington}}, \bibinfo {author}
  {\bibfnamefont {O.~A.}\ \bibnamefont {Hurricane}}, \bibinfo {author}
  {\bibfnamefont {A.~M.}\ \bibnamefont {Dimits}}, \ and\ \bibinfo {author}
  {\bibfnamefont {B.~A.}\ \bibnamefont {Remington}},\ }\bibfield  {title}
  {\enquote {\bibinfo {title} {Turbulent mixing and transition criteria of
  flows induced by hydrodynamic instabilities},}\ }\href {\doibase {}}
  {\bibfield  {journal} {\bibinfo  {journal} {Physics of Plasmas}\ }\textbf
  {\bibinfo {volume} {26}},\ \bibinfo {pages} {080901} (\bibinfo {year}
  {2019})}\BibitemShut {NoStop}%
\bibitem [{\citenamefont {Waddell}, \citenamefont {Niederhaus},\ and\
  \citenamefont {Jacobs}(2001)}]{Waddell_2001}%
  \BibitemOpen
  \bibfield  {author} {\bibinfo {author} {\bibfnamefont {J.}~\bibnamefont
  {Waddell}}, \bibinfo {author} {\bibfnamefont {C.}~\bibnamefont {Niederhaus}},
  \ and\ \bibinfo {author} {\bibfnamefont {J.~W.}\ \bibnamefont {Jacobs}},\
  }\bibfield  {title} {\enquote {\bibinfo {title} {Experimental study of
  {Rayleigh--Taylor} instability: low {Atwood} number liquid systems with
  single-mode initial perturbations},}\ }\href {\doibase {}} {\bibfield
  {journal} {\bibinfo  {journal} {Physics of Fluids}\ }\textbf {\bibinfo
  {volume} {13}},\ \bibinfo {pages} {1263--1273} (\bibinfo {year}
  {2001})}\BibitemShut {NoStop}%
\bibitem [{\citenamefont {Wilkinson}\ and\ \citenamefont
  {Jacobs}(2007)}]{Wilkinson_2007}%
  \BibitemOpen
  \bibfield  {author} {\bibinfo {author} {\bibfnamefont {J.}~\bibnamefont
  {Wilkinson}}\ and\ \bibinfo {author} {\bibfnamefont {J.~W.}\ \bibnamefont
  {Jacobs}},\ }\bibfield  {title} {\enquote {\bibinfo {title} {Experimental
  study of the single-mode three-dimensional {Rayleigh--Taylor} instability},}\
  }\href {\doibase {}} {\bibfield  {journal} {\bibinfo  {journal} {Physics of
  fluids}\ }\textbf {\bibinfo {volume} {19}},\ \bibinfo {pages} {124102}
  (\bibinfo {year} {2007})}\BibitemShut {NoStop}%
\bibitem [{\citenamefont {Luo}\ \emph {et~al.}(2018)\citenamefont {Luo},
  \citenamefont {Zhang}, \citenamefont {Ding}, \citenamefont {Si},
  \citenamefont {Yang}, \citenamefont {Zhai},\ and\ \citenamefont
  {Wen}}]{Luoxs_2018}%
  \BibitemOpen
  \bibfield  {author} {\bibinfo {author} {\bibfnamefont {X.}~\bibnamefont
  {Luo}}, \bibinfo {author} {\bibfnamefont {F.}~\bibnamefont {Zhang}}, \bibinfo
  {author} {\bibfnamefont {J.}~\bibnamefont {Ding}}, \bibinfo {author}
  {\bibfnamefont {T.}~\bibnamefont {Si}}, \bibinfo {author} {\bibfnamefont
  {J.}~\bibnamefont {Yang}}, \bibinfo {author} {\bibfnamefont {Z.}~\bibnamefont
  {Zhai}}, \ and\ \bibinfo {author} {\bibfnamefont {C.~Y.}\ \bibnamefont
  {Wen}},\ }\bibfield  {title} {\enquote {\bibinfo {title} {Long-term effect of
  {Rayleigh--Taylor} stabilization on converging {Richtmyer--Meshkov}
  instability},}\ }\href {\doibase {}} {\bibfield  {journal} {\bibinfo
  {journal} {Journal of Fluid Mechanics}\ }\textbf {\bibinfo {volume} {849}},\
  \bibinfo {pages} {231--244} (\bibinfo {year} {2018})}\BibitemShut {NoStop}%
\bibitem [{\citenamefont {Chang}\ \emph {et~al.}(1996)\citenamefont {Chang},
  \citenamefont {Hou}, \citenamefont {Merriman},\ and\ \citenamefont
  {Osher}}]{Chang_1996}%
  \BibitemOpen
  \bibfield  {author} {\bibinfo {author} {\bibfnamefont {Y.-C.}\ \bibnamefont
  {Chang}}, \bibinfo {author} {\bibfnamefont {T.}~\bibnamefont {Hou}}, \bibinfo
  {author} {\bibfnamefont {B.}~\bibnamefont {Merriman}}, \ and\ \bibinfo
  {author} {\bibfnamefont {S.}~\bibnamefont {Osher}},\ }\bibfield  {title}
  {\enquote {\bibinfo {title} {A level set formulation of {Eulerian} interface
  capturing methods for incompressible fluid flows},}\ }\href {\doibase {}}
  {\bibfield  {journal} {\bibinfo  {journal} {Journal of Computational
  Physics}\ }\textbf {\bibinfo {volume} {124}},\ \bibinfo {pages} {449--464}
  (\bibinfo {year} {1996})}\BibitemShut {NoStop}%
\bibitem [{\citenamefont {Glimm}\ \emph {et~al.}(1986)\citenamefont {Glimm},
  \citenamefont {McBryan}, \citenamefont {Menikoff},\ and\ \citenamefont
  {Sharp}}]{Glimm_1986}%
  \BibitemOpen
  \bibfield  {author} {\bibinfo {author} {\bibfnamefont {J.}~\bibnamefont
  {Glimm}}, \bibinfo {author} {\bibfnamefont {O.}~\bibnamefont {McBryan}},
  \bibinfo {author} {\bibfnamefont {R.}~\bibnamefont {Menikoff}}, \ and\
  \bibinfo {author} {\bibfnamefont {D.}~\bibnamefont {Sharp}},\ }\bibfield
  {title} {\enquote {\bibinfo {title} {Front tracking applied to
  {Rayleigh--Taylor} instability},}\ }\href {\doibase {}} {\bibfield  {journal}
  {\bibinfo  {journal} {SIAM Journal on Scientific and Statistical Computing}\
  }\textbf {\bibinfo {volume} {7}},\ \bibinfo {pages} {230--251} (\bibinfo
  {year} {1986})}\BibitemShut {NoStop}%
\bibitem [{\citenamefont {Gerlach}\ \emph {et~al.}(2006)\citenamefont
  {Gerlach}, \citenamefont {Tomar}, \citenamefont {Biswas},\ and\ \citenamefont
  {Durst}}]{Gerlach_2006}%
  \BibitemOpen
  \bibfield  {author} {\bibinfo {author} {\bibfnamefont {D.}~\bibnamefont
  {Gerlach}}, \bibinfo {author} {\bibfnamefont {G.}~\bibnamefont {Tomar}},
  \bibinfo {author} {\bibfnamefont {G.}~\bibnamefont {Biswas}}, \ and\ \bibinfo
  {author} {\bibfnamefont {F.}~\bibnamefont {Durst}},\ }\bibfield  {title}
  {\enquote {\bibinfo {title} {Comparison of volume-of-fluid methods for
  surface tension-dominant two-phase flows},}\ }\href {\doibase {}} {\bibfield
  {journal} {\bibinfo  {journal} {International Journal of Heat and Mass
  Transfer}\ }\textbf {\bibinfo {volume} {49}},\ \bibinfo {pages} {740--754}
  (\bibinfo {year} {2006})}\BibitemShut {NoStop}%
\bibitem [{\citenamefont {Shadloo}, \citenamefont {Zainali},\ and\
  \citenamefont {Yildiz}(2013)}]{Shadloo_2013}%
  \BibitemOpen
  \bibfield  {author} {\bibinfo {author} {\bibfnamefont {M.}~\bibnamefont
  {Shadloo}}, \bibinfo {author} {\bibfnamefont {A.}~\bibnamefont {Zainali}}, \
  and\ \bibinfo {author} {\bibfnamefont {M.}~\bibnamefont {Yildiz}},\
  }\bibfield  {title} {\enquote {\bibinfo {title} {Simulation of single mode
  {Rayleigh--Taylor} instability by {SPH} method},}\ }\href {\doibase {}}
  {\bibfield  {journal} {\bibinfo  {journal} {Computational Mechanics}\
  }\textbf {\bibinfo {volume} {51}},\ \bibinfo {pages} {699--715} (\bibinfo
  {year} {2013})}\BibitemShut {NoStop}%
\bibitem [{\citenamefont {Mellado}, \citenamefont {Sarkar},\ and\ \citenamefont
  {Zhou}(2005)}]{Mellado_2005}%
  \BibitemOpen
  \bibfield  {author} {\bibinfo {author} {\bibfnamefont {J.~P.}\ \bibnamefont
  {Mellado}}, \bibinfo {author} {\bibfnamefont {S.}~\bibnamefont {Sarkar}}, \
  and\ \bibinfo {author} {\bibfnamefont {Y.}~\bibnamefont {Zhou}},\ }\bibfield
  {title} {\enquote {\bibinfo {title} {Large-eddy simulation of
  {Rayleigh--Taylor} turbulence with compressible miscible fluids},}\ }\href
  {\doibase {}} {\bibfield  {journal} {\bibinfo  {journal} {Physics of Fluids}\
  }\textbf {\bibinfo {volume} {17}},\ \bibinfo {pages} {076101} (\bibinfo
  {year} {2005})}\BibitemShut {NoStop}%
\bibitem [{\citenamefont {Celani}\ \emph {et~al.}(2009)\citenamefont {Celani},
  \citenamefont {Mazzino}, \citenamefont {Muratore-Ginanneschi},\ and\
  \citenamefont {Vozella}}]{Celani_2009}%
  \BibitemOpen
  \bibfield  {author} {\bibinfo {author} {\bibfnamefont {A.}~\bibnamefont
  {Celani}}, \bibinfo {author} {\bibfnamefont {A.}~\bibnamefont {Mazzino}},
  \bibinfo {author} {\bibfnamefont {P.}~\bibnamefont {Muratore-Ginanneschi}}, \
  and\ \bibinfo {author} {\bibfnamefont {L.}~\bibnamefont {Vozella}},\
  }\bibfield  {title} {\enquote {\bibinfo {title} {Phase-field model for the
  {Rayleigh--Taylor} instability of immiscible fluids},}\ }\href {\doibase {}}
  {\bibfield  {journal} {\bibinfo  {journal} {Journal of Fluid Mechanics}\
  }\textbf {\bibinfo {volume} {622}},\ \bibinfo {pages} {115--134} (\bibinfo
  {year} {2009})}\BibitemShut {NoStop}%
\bibitem [{\citenamefont {Mueschke}\ and\ \citenamefont
  {Schilling}(2009)}]{Mueschke_2009}%
  \BibitemOpen
  \bibfield  {author} {\bibinfo {author} {\bibfnamefont {N.~J.}\ \bibnamefont
  {Mueschke}}\ and\ \bibinfo {author} {\bibfnamefont {O.}~\bibnamefont
  {Schilling}},\ }\bibfield  {title} {\enquote {\bibinfo {title} {Investigation
  of {Rayleigh--Taylor} turbulence and mixing using direct numerical simulation
  with experimentally measured initial conditions. {I}. {Comparison} to
  experimental data},}\ }\href {\doibase {}} {\bibfield  {journal} {\bibinfo
  {journal} {Physics of Fluids}\ }\textbf {\bibinfo {volume} {21}},\ \bibinfo
  {pages} {014106} (\bibinfo {year} {2009})}\BibitemShut {NoStop}%
\bibitem [{\citenamefont {Young}\ \emph {et~al.}(2001)\citenamefont {Young},
  \citenamefont {Tufo}, \citenamefont {Dubey},\ and\ \citenamefont
  {Rosner}}]{Young_2001}%
  \BibitemOpen
  \bibfield  {author} {\bibinfo {author} {\bibfnamefont {Y.-N.}\ \bibnamefont
  {Young}}, \bibinfo {author} {\bibfnamefont {H.}~\bibnamefont {Tufo}},
  \bibinfo {author} {\bibfnamefont {A.}~\bibnamefont {Dubey}}, \ and\ \bibinfo
  {author} {\bibfnamefont {R.}~\bibnamefont {Rosner}},\ }\bibfield  {title}
  {\enquote {\bibinfo {title} {On the miscible {Rayleigh--Taylor} instability:
  two and three dimensions},}\ }\href {\doibase {}} {\bibfield  {journal}
  {\bibinfo  {journal} {Journal of Fluid Mechanics}\ }\textbf {\bibinfo
  {volume} {447}},\ \bibinfo {pages} {377--408} (\bibinfo {year}
  {2001})}\BibitemShut {NoStop}%
\bibitem [{\citenamefont {Wei}\ and\ \citenamefont {Livescu}(2012)}]{Wei_2012}%
  \BibitemOpen
  \bibfield  {author} {\bibinfo {author} {\bibfnamefont {T.}~\bibnamefont
  {Wei}}\ and\ \bibinfo {author} {\bibfnamefont {D.}~\bibnamefont {Livescu}},\
  }\bibfield  {title} {\enquote {\bibinfo {title} {Late-time quadratic growth
  in single-mode {Rayleigh--Taylor} instability},}\ }\href {\doibase {}}
  {\bibfield  {journal} {\bibinfo  {journal} {Physical Review E}\ }\textbf
  {\bibinfo {volume} {86}},\ \bibinfo {pages} {046405} (\bibinfo {year}
  {2012})}\BibitemShut {NoStop}%
\bibitem [{\citenamefont {Liang}\ \emph {et~al.}(2019)\citenamefont {Liang},
  \citenamefont {Hu}, \citenamefont {Huang},\ and\ \citenamefont
  {Xu}}]{Liang_2019}%
  \BibitemOpen
  \bibfield  {author} {\bibinfo {author} {\bibfnamefont {H.}~\bibnamefont
  {Liang}}, \bibinfo {author} {\bibfnamefont {X.}~\bibnamefont {Hu}}, \bibinfo
  {author} {\bibfnamefont {X.}~\bibnamefont {Huang}}, \ and\ \bibinfo {author}
  {\bibfnamefont {J.}~\bibnamefont {Xu}},\ }\bibfield  {title} {\enquote
  {\bibinfo {title} {Direct numerical simulations of multi-mode immiscible
  {Rayleigh--Taylor} instability with high {Reynolds} numbers},}\ }\href
  {\doibase {}} {\bibfield  {journal} {\bibinfo  {journal} {Physics of Fluids}\
  }\textbf {\bibinfo {volume} {31}},\ \bibinfo {pages} {112104} (\bibinfo
  {year} {2019})}\BibitemShut {NoStop}%
\bibitem [{\citenamefont {Xie}\ \emph {et~al.}(2017)\citenamefont {Xie},
  \citenamefont {Tao}, \citenamefont {Sun},\ and\ \citenamefont
  {Li}}]{Xie_2017}%
  \BibitemOpen
  \bibfield  {author} {\bibinfo {author} {\bibfnamefont {C.~Y.}\ \bibnamefont
  {Xie}}, \bibinfo {author} {\bibfnamefont {J.~J.}\ \bibnamefont {Tao}},
  \bibinfo {author} {\bibfnamefont {Z.~L.}\ \bibnamefont {Sun}}, \ and\
  \bibinfo {author} {\bibfnamefont {J.}~\bibnamefont {Li}},\ }\bibfield
  {title} {\enquote {\bibinfo {title} {Retarding viscous {Rayleigh--Taylor}
  mixing by an optimized additional mode},}\ }\href {\doibase {}} {\bibfield
  {journal} {\bibinfo  {journal} {Physical Review E}\ }\textbf {\bibinfo
  {volume} {95}},\ \bibinfo {pages} {023109} (\bibinfo {year}
  {2017})}\BibitemShut {NoStop}%
\bibitem [{\citenamefont {Hu}\ \emph {et~al.}(2019)\citenamefont {Hu},
  \citenamefont {Zhang}, \citenamefont {Tian}, \citenamefont {He},\ and\
  \citenamefont {Li}}]{Hu_pof_2019}%
  \BibitemOpen
  \bibfield  {author} {\bibinfo {author} {\bibfnamefont {Z.-X.}\ \bibnamefont
  {Hu}}, \bibinfo {author} {\bibfnamefont {Y.-S.}\ \bibnamefont {Zhang}},
  \bibinfo {author} {\bibfnamefont {B.}~\bibnamefont {Tian}}, \bibinfo {author}
  {\bibfnamefont {Z.}~\bibnamefont {He}}, \ and\ \bibinfo {author}
  {\bibfnamefont {L.}~\bibnamefont {Li}},\ }\bibfield  {title} {\enquote
  {\bibinfo {title} {Effect of viscosity on two-dimensional single-mode
  {Rayleigh--Taylor} instability during and after the reacceleration stage},}\
  }\href {\doibase {}} {\bibfield  {journal} {\bibinfo  {journal} {Physics of
  Fluids}\ }\textbf {\bibinfo {volume} {31}},\ \bibinfo {pages} {104108}
  (\bibinfo {year} {2019})}\BibitemShut {NoStop}%
\bibitem [{\citenamefont {Xu}, \citenamefont {Zhang},\ and\ \citenamefont
  {Zhang}(2018)}]{Xu_2018}%
  \BibitemOpen
  \bibfield  {author} {\bibinfo {author} {\bibfnamefont {A.}~\bibnamefont
  {Xu}}, \bibinfo {author} {\bibfnamefont {G.}~\bibnamefont {Zhang}}, \ and\
  \bibinfo {author} {\bibfnamefont {Y.}~\bibnamefont {Zhang}},\ }\bibfield
  {title} {\enquote {\bibinfo {title} {Discrete {Boltzmann} modeling of
  compressible flows},}\ }in\ \href {\doibase {}} {\emph {\bibinfo {booktitle}
  {Kinetic Theory}}}\ (\bibinfo  {publisher} {IntecOpen},\ \bibinfo {year}
  {2018})\ pp.\ \bibinfo {pages} {450--458}\BibitemShut {NoStop}%
\bibitem [{\citenamefont {Gan}\ \emph {et~al.}(2019)\citenamefont {Gan},
  \citenamefont {Xu}, \citenamefont {Zhang}, \citenamefont {Lin}, \citenamefont
  {Lai},\ and\ \citenamefont {Liu}}]{Gan_2019}%
  \BibitemOpen
  \bibfield  {author} {\bibinfo {author} {\bibfnamefont {Y.~B.}\ \bibnamefont
  {Gan}}, \bibinfo {author} {\bibfnamefont {A.~G.}\ \bibnamefont {Xu}},
  \bibinfo {author} {\bibfnamefont {G.~C.}\ \bibnamefont {Zhang}}, \bibinfo
  {author} {\bibfnamefont {C.~D.}\ \bibnamefont {Lin}}, \bibinfo {author}
  {\bibfnamefont {H.~L.}\ \bibnamefont {Lai}}, \ and\ \bibinfo {author}
  {\bibfnamefont {Z.~P.}\ \bibnamefont {Liu}},\ }\bibfield  {title} {\enquote
  {\bibinfo {title} {Nonequilibrium and morphological characterizations of
  {Kelvin--Helmholtz} instability in compressible flows},}\ }\href {\doibase
  {}} {\bibfield  {journal} {\bibinfo  {journal} {Frontiers of Physics}\
  }\textbf {\bibinfo {volume} {14}},\ \bibinfo {pages} {1--17} (\bibinfo {year}
  {2019})}\BibitemShut {NoStop}%
\bibitem [{\citenamefont {Lai}\ \emph {et~al.}(2016)\citenamefont {Lai},
  \citenamefont {Xu}, \citenamefont {Zhang}, \citenamefont {Gan}, \citenamefont
  {Ying},\ and\ \citenamefont {Succi}}]{Lai_2016}%
  \BibitemOpen
  \bibfield  {author} {\bibinfo {author} {\bibfnamefont {H.}~\bibnamefont
  {Lai}}, \bibinfo {author} {\bibfnamefont {A.}~\bibnamefont {Xu}}, \bibinfo
  {author} {\bibfnamefont {G.}~\bibnamefont {Zhang}}, \bibinfo {author}
  {\bibfnamefont {Y.}~\bibnamefont {Gan}}, \bibinfo {author} {\bibfnamefont
  {Y.}~\bibnamefont {Ying}}, \ and\ \bibinfo {author} {\bibfnamefont
  {S.}~\bibnamefont {Succi}},\ }\bibfield  {title} {\enquote {\bibinfo {title}
  {Nonequilibrium thermohydrodynamic effects on the {Rayleigh--Taylor}
  instability in compressible flows},}\ }\href {\doibase {}} {\bibfield
  {journal} {\bibinfo  {journal} {Physical Review E}\ }\textbf {\bibinfo
  {volume} {94}},\ \bibinfo {pages} {023106} (\bibinfo {year}
  {2016})}\BibitemShut {NoStop}%
\bibitem [{\citenamefont {Chen}, \citenamefont {Xu},\ and\ \citenamefont
  {Zhang}(2016)}]{Chen_2016}%
  \BibitemOpen
  \bibfield  {author} {\bibinfo {author} {\bibfnamefont {F.}~\bibnamefont
  {Chen}}, \bibinfo {author} {\bibfnamefont {A.}~\bibnamefont {Xu}}, \ and\
  \bibinfo {author} {\bibfnamefont {G.}~\bibnamefont {Zhang}},\ }\bibfield
  {title} {\enquote {\bibinfo {title} {Viscosity, heat conductivity, and
  {Prandtl} number effects in the {Rayleigh--Taylor} instability},}\ }\href
  {\doibase {}} {\bibfield  {journal} {\bibinfo  {journal} {Frontiers of
  Physics}\ }\textbf {\bibinfo {volume} {11}},\ \bibinfo {pages} {114703}
  (\bibinfo {year} {2016})}\BibitemShut {NoStop}%
\bibitem [{\citenamefont {Chen}, \citenamefont {Xu},\ and\ \citenamefont
  {Zhang}(2018)}]{Chen_2018}%
  \BibitemOpen
  \bibfield  {author} {\bibinfo {author} {\bibfnamefont {F.}~\bibnamefont
  {Chen}}, \bibinfo {author} {\bibfnamefont {A.}~\bibnamefont {Xu}}, \ and\
  \bibinfo {author} {\bibfnamefont {G.}~\bibnamefont {Zhang}},\ }\bibfield
  {title} {\enquote {\bibinfo {title} {Collaboration and competition between
  {Richtmyer--Meshkov} instability and {Rayleigh--Taylor} instability},}\
  }\href {\doibase {}} {\bibfield  {journal} {\bibinfo  {journal} {Physics of
  Fluids}\ }\textbf {\bibinfo {volume} {30}},\ \bibinfo {pages} {102105}
  (\bibinfo {year} {2018})}\BibitemShut {NoStop}%
\bibitem [{\citenamefont {Lin}\ \emph {et~al.}(2017)\citenamefont {Lin},
  \citenamefont {Xu}, \citenamefont {Zhang}, \citenamefont {Luo},\ and\
  \citenamefont {Li}}]{Lin_2017}%
  \BibitemOpen
  \bibfield  {author} {\bibinfo {author} {\bibfnamefont {C.}~\bibnamefont
  {Lin}}, \bibinfo {author} {\bibfnamefont {A.}~\bibnamefont {Xu}}, \bibinfo
  {author} {\bibfnamefont {G.}~\bibnamefont {Zhang}}, \bibinfo {author}
  {\bibfnamefont {K.~H.}\ \bibnamefont {Luo}}, \ and\ \bibinfo {author}
  {\bibfnamefont {Y.}~\bibnamefont {Li}},\ }\bibfield  {title} {\enquote
  {\bibinfo {title} {Discrete {Boltzmann} modeling of {Rayleigh--Taylor}
  instability in two-component compressible flows},}\ }\href {\doibase {}}
  {\bibfield  {journal} {\bibinfo  {journal} {Physical Review E}\ }\textbf
  {\bibinfo {volume} {96}},\ \bibinfo {pages} {053305} (\bibinfo {year}
  {2017})}\BibitemShut {NoStop}%
\bibitem [{\citenamefont {Zhang}\ \emph {et~al.}(2019)\citenamefont {Zhang},
  \citenamefont {Xu}, \citenamefont {Zhang}, \citenamefont {Chen},\ and\
  \citenamefont {Pei}}]{ZhangYD_2019}%
  \BibitemOpen
  \bibfield  {author} {\bibinfo {author} {\bibfnamefont {Y.}~\bibnamefont
  {Zhang}}, \bibinfo {author} {\bibfnamefont {A.}~\bibnamefont {Xu}}, \bibinfo
  {author} {\bibfnamefont {G.}~\bibnamefont {Zhang}}, \bibinfo {author}
  {\bibfnamefont {Z.}~\bibnamefont {Chen}}, \ and\ \bibinfo {author}
  {\bibfnamefont {W.}~\bibnamefont {Pei}},\ }\bibfield  {title} {\enquote
  {\bibinfo {title} {Discrete {Boltzmann} method for non-equilibrium flows:
  based on {Shakhov} model},}\ }\href {\doibase {}} {\bibfield  {journal}
  {\bibinfo  {journal} {Computer Physics Communications}\ }\textbf {\bibinfo
  {volume} {238}},\ \bibinfo {pages} {50--65} (\bibinfo {year}
  {2019})}\BibitemShut {NoStop}%
\bibitem [{\citenamefont {Boffetta}, \citenamefont {Borgnino},\ and\
  \citenamefont {Musacchio}(2020)}]{boffetta2020scaling}%
  \BibitemOpen
  \bibfield  {author} {\bibinfo {author} {\bibfnamefont {G.}~\bibnamefont
  {Boffetta}}, \bibinfo {author} {\bibfnamefont {M.}~\bibnamefont {Borgnino}},
  \ and\ \bibinfo {author} {\bibfnamefont {S.}~\bibnamefont {Musacchio}},\
  }\bibfield  {title} {\enquote {\bibinfo {title} {Scaling of {Rayleigh-Taylor}
  mixing in porous media},}\ }\href@noop {} {\bibfield  {journal} {\bibinfo
  {journal} {Physical Review Fluids}\ }\textbf {\bibinfo {volume} {5}},\
  \bibinfo {pages} {062501} (\bibinfo {year} {2020})}\BibitemShut {NoStop}%
\bibitem [{\citenamefont {Cook}\ and\ \citenamefont {Zhou}(2002)}]{Cook_2002}%
  \BibitemOpen
  \bibfield  {author} {\bibinfo {author} {\bibfnamefont {A.~W.}\ \bibnamefont
  {Cook}}\ and\ \bibinfo {author} {\bibfnamefont {Y.}~\bibnamefont {Zhou}},\
  }\bibfield  {title} {\enquote {\bibinfo {title} {Energy transfer in
  {Rayleigh--Taylor} instability},}\ }\href {\doibase {}} {\bibfield  {journal}
  {\bibinfo  {journal} {Physical Review E}\ }\textbf {\bibinfo {volume} {66}},\
  \bibinfo {pages} {026312} (\bibinfo {year} {2002})}\BibitemShut {NoStop}%
\bibitem [{\citenamefont {Cabot}\ and\ \citenamefont
  {Cook}(2006)}]{Cabot_2006}%
  \BibitemOpen
  \bibfield  {author} {\bibinfo {author} {\bibfnamefont {W.~H.}\ \bibnamefont
  {Cabot}}\ and\ \bibinfo {author} {\bibfnamefont {A.~W.}\ \bibnamefont
  {Cook}},\ }\bibfield  {title} {\enquote {\bibinfo {title} {{Reynolds} number
  effects on {Rayleigh--Taylor} instability with possible implications for type
  {Ia} supernovae},}\ }\href {\doibase {}} {\bibfield  {journal} {\bibinfo
  {journal} {Nature Physics}\ }\textbf {\bibinfo {volume} {2}},\ \bibinfo
  {pages} {562} (\bibinfo {year} {2006})}\BibitemShut {NoStop}%
\bibitem [{\citenamefont {Olson}\ and\ \citenamefont
  {Jacobs}(2009)}]{Olson_2009}%
  \BibitemOpen
  \bibfield  {author} {\bibinfo {author} {\bibfnamefont {D.}~\bibnamefont
  {Olson}}\ and\ \bibinfo {author} {\bibfnamefont {J.}~\bibnamefont {Jacobs}},\
  }\bibfield  {title} {\enquote {\bibinfo {title} {Experimental study of
  {Rayleigh--Taylor} instability with a complex initial perturbation},}\ }\href
  {\doibase {}} {\bibfield  {journal} {\bibinfo  {journal} {Physics of Fluids}\
  }\textbf {\bibinfo {volume} {21}},\ \bibinfo {pages} {034103} (\bibinfo
  {year} {2009})}\BibitemShut {NoStop}%
\bibitem [{\citenamefont {Dimonte}\ \emph {et~al.}(2004)\citenamefont
  {Dimonte}, \citenamefont {Youngs}, \citenamefont {Dimits}, \citenamefont
  {Weber}, \citenamefont {Marinak}, \citenamefont {Wunsch}, \citenamefont
  {Garasi}, \citenamefont {Robinson}, \citenamefont {Andrews},\ and\
  \citenamefont {Ramaprabhu}}]{Dimonte_2004}%
  \BibitemOpen
  \bibfield  {author} {\bibinfo {author} {\bibfnamefont {G.}~\bibnamefont
  {Dimonte}}, \bibinfo {author} {\bibfnamefont {D.}~\bibnamefont {Youngs}},
  \bibinfo {author} {\bibfnamefont {A.}~\bibnamefont {Dimits}}, \bibinfo
  {author} {\bibfnamefont {S.}~\bibnamefont {Weber}}, \bibinfo {author}
  {\bibfnamefont {M.}~\bibnamefont {Marinak}}, \bibinfo {author} {\bibfnamefont
  {S.}~\bibnamefont {Wunsch}}, \bibinfo {author} {\bibfnamefont
  {C.}~\bibnamefont {Garasi}}, \bibinfo {author} {\bibfnamefont
  {A.}~\bibnamefont {Robinson}}, \bibinfo {author} {\bibfnamefont
  {M.}~\bibnamefont {Andrews}}, \ and\ \bibinfo {author} {\bibfnamefont
  {P.}~\bibnamefont {Ramaprabhu}},\ }\bibfield  {title} {\enquote {\bibinfo
  {title} {A comparative study of the turbulent {Rayleigh--Taylor} instability
  using high-resolution three-dimensional numerical simulations: the
  {Alpha--Group} collaboration},}\ }\href {\doibase {}} {\bibfield  {journal}
  {\bibinfo  {journal} {Physics of Fluids}\ }\textbf {\bibinfo {volume} {16}},\
  \bibinfo {pages} {1668--1693} (\bibinfo {year} {2004})}\BibitemShut {NoStop}%
\bibitem [{\citenamefont {Zhou}\ and\ \citenamefont
  {Cabot}(2019)}]{ZhouY_pof_2019}%
  \BibitemOpen
  \bibfield  {author} {\bibinfo {author} {\bibfnamefont {Y.}~\bibnamefont
  {Zhou}}\ and\ \bibinfo {author} {\bibfnamefont {W.~H.}\ \bibnamefont
  {Cabot}},\ }\bibfield  {title} {\enquote {\bibinfo {title} {Time-dependent
  study of anisotropy in {Rayleigh--Taylor} instability induced turbulent flows
  with a variety of density ratios},}\ }\href {\doibase {}} {\bibfield
  {journal} {\bibinfo  {journal} {Physics of Fluids}\ }\textbf {\bibinfo
  {volume} {31}},\ \bibinfo {pages} {084106} (\bibinfo {year}
  {2019})}\BibitemShut {NoStop}%
\bibitem [{\citenamefont {Zhou}, \citenamefont {Cabot},\ and\ \citenamefont
  {Thornber}(2016)}]{ZhouY_pop_2019}%
  \BibitemOpen
  \bibfield  {author} {\bibinfo {author} {\bibfnamefont {Y.}~\bibnamefont
  {Zhou}}, \bibinfo {author} {\bibfnamefont {W.~H.}\ \bibnamefont {Cabot}}, \
  and\ \bibinfo {author} {\bibfnamefont {B.}~\bibnamefont {Thornber}},\
  }\bibfield  {title} {\enquote {\bibinfo {title} {Asymptotic behavior of the
  mixed mass in {Rayleigh--Taylor} and {Richtmyer--Meshkov} instability induced
  flows},}\ }\href@noop {} {\bibfield  {journal} {\bibinfo  {journal} {Physics
  of Plasmas}\ }\textbf {\bibinfo {volume} {23}},\ \bibinfo {pages} {052712}
  (\bibinfo {year} {2016})}\BibitemShut {NoStop}%
\bibitem [{\citenamefont {Ma}\ \emph {et~al.}(2017)\citenamefont {Ma},
  \citenamefont {Patel}, \citenamefont {Izumi}, \citenamefont {Springer},
  \citenamefont {Key}, \citenamefont {Atherton}, \citenamefont {Barrios},
  \citenamefont {Benedetti}, \citenamefont {Bionta}, \citenamefont {Bond} \emph
  {et~al.}}]{MaT_pop_2017}%
  \BibitemOpen
  \bibfield  {author} {\bibinfo {author} {\bibfnamefont {T.}~\bibnamefont
  {Ma}}, \bibinfo {author} {\bibfnamefont {P.}~\bibnamefont {Patel}}, \bibinfo
  {author} {\bibfnamefont {N.}~\bibnamefont {Izumi}}, \bibinfo {author}
  {\bibfnamefont {P.}~\bibnamefont {Springer}}, \bibinfo {author}
  {\bibfnamefont {M.}~\bibnamefont {Key}}, \bibinfo {author} {\bibfnamefont
  {L.}~\bibnamefont {Atherton}}, \bibinfo {author} {\bibfnamefont
  {M.}~\bibnamefont {Barrios}}, \bibinfo {author} {\bibfnamefont
  {L.}~\bibnamefont {Benedetti}}, \bibinfo {author} {\bibfnamefont
  {R.}~\bibnamefont {Bionta}}, \bibinfo {author} {\bibfnamefont
  {E.}~\bibnamefont {Bond}},  \emph {et~al.},\ }\bibfield  {title} {\enquote
  {\bibinfo {title} {The role of hot spot mix in the low-foot and high-foot
  implosions on the {NIF}},}\ }\href@noop {} {\bibfield  {journal} {\bibinfo
  {journal} {Physics of Plasmas}\ }\textbf {\bibinfo {volume} {24}},\ \bibinfo
  {pages} {056311} (\bibinfo {year} {2017})}\BibitemShut {NoStop}%
\bibitem [{\citenamefont {Gan}\ \emph {et~al.}(2013)\citenamefont {Gan},
  \citenamefont {Xu}, \citenamefont {Zhang},\ and\ \citenamefont
  {Yang}}]{Gan_2013}%
  \BibitemOpen
  \bibfield  {author} {\bibinfo {author} {\bibfnamefont {Y.}~\bibnamefont
  {Gan}}, \bibinfo {author} {\bibfnamefont {A.}~\bibnamefont {Xu}}, \bibinfo
  {author} {\bibfnamefont {G.}~\bibnamefont {Zhang}}, \ and\ \bibinfo {author}
  {\bibfnamefont {Y.}~\bibnamefont {Yang}},\ }\bibfield  {title} {\enquote
  {\bibinfo {title} {Lattice {BGK} kinetic model for high-speed compressible
  flows: {Hydrodynamic} and nonequilibrium behaviors},}\ }\href {\doibase {}}
  {\bibfield  {journal} {\bibinfo  {journal} {EPL (Europhysics Letters)}\
  }\textbf {\bibinfo {volume} {103}},\ \bibinfo {pages} {24003} (\bibinfo
  {year} {2013})}\BibitemShut {NoStop}%
\bibitem [{\citenamefont {Bhatnagar}, \citenamefont {Gross},\ and\
  \citenamefont {Krook}(1954)}]{Bhatnagar_1954}%
  \BibitemOpen
  \bibfield  {author} {\bibinfo {author} {\bibfnamefont {P.~L.}\ \bibnamefont
  {Bhatnagar}}, \bibinfo {author} {\bibfnamefont {E.~P.}\ \bibnamefont
  {Gross}}, \ and\ \bibinfo {author} {\bibfnamefont {M.~K.}\ \bibnamefont
  {Krook}},\ }\bibfield  {title} {\enquote {\bibinfo {title} {A model for
  collision processes in gases. {I}. {Small} amplitude processes in charged and
  neutral one-component systems},}\ }\href {\doibase {}} {\bibfield  {journal}
  {\bibinfo  {journal} {Physical Review}\ }\textbf {\bibinfo {volume} {94}},\
  \bibinfo {pages} {511} (\bibinfo {year} {1954})}\BibitemShut {NoStop}%
\bibitem [{\citenamefont {He}, \citenamefont {Shan},\ and\ \citenamefont
  {Doolen}(1998)}]{HeShan_1998}%
  \BibitemOpen
  \bibfield  {author} {\bibinfo {author} {\bibfnamefont {X.}~\bibnamefont
  {He}}, \bibinfo {author} {\bibfnamefont {X.}~\bibnamefont {Shan}}, \ and\
  \bibinfo {author} {\bibfnamefont {G.~D.}\ \bibnamefont {Doolen}},\ }\bibfield
   {title} {\enquote {\bibinfo {title} {Discrete {Boltzmann} equation model for
  nonideal gases},}\ }\href {\doibase {}} {\bibfield  {journal} {\bibinfo
  {journal} {Physical Review E}\ }\textbf {\bibinfo {volume} {57}},\ \bibinfo
  {pages} {R13} (\bibinfo {year} {1998})}\BibitemShut {NoStop}%
\bibitem [{\citenamefont {Xu}\ \emph {et~al.}(2012)\citenamefont {Xu},
  \citenamefont {Zhang}, \citenamefont {Gan}, \citenamefont {Chen},\ and\
  \citenamefont {Yu}}]{XuAG_2012}%
  \BibitemOpen
  \bibfield  {author} {\bibinfo {author} {\bibfnamefont {A.}~\bibnamefont
  {Xu}}, \bibinfo {author} {\bibfnamefont {G.}~\bibnamefont {Zhang}}, \bibinfo
  {author} {\bibfnamefont {Y.}~\bibnamefont {Gan}}, \bibinfo {author}
  {\bibfnamefont {F.}~\bibnamefont {Chen}}, \ and\ \bibinfo {author}
  {\bibfnamefont {X.}~\bibnamefont {Yu}},\ }\bibfield  {title} {\enquote
  {\bibinfo {title} {Lattice {Boltzmann} modeling and simulation of
  compressible flows},}\ }\href {\doibase {}} {\bibfield  {journal} {\bibinfo
  {journal} {Frontiers of Physics}\ }\textbf {\bibinfo {volume} {7}},\ \bibinfo
  {pages} {582--600} (\bibinfo {year} {2012})}\BibitemShut {NoStop}%
\bibitem [{\citenamefont {Xu}\ \emph {et~al.}(2015)\citenamefont {Xu},
  \citenamefont {Lin}, \citenamefont {Zhang},\ and\ \citenamefont
  {Li}}]{xu2015multiple}%
  \BibitemOpen
  \bibfield  {author} {\bibinfo {author} {\bibfnamefont {A.}~\bibnamefont
  {Xu}}, \bibinfo {author} {\bibfnamefont {C.}~\bibnamefont {Lin}}, \bibinfo
  {author} {\bibfnamefont {G.}~\bibnamefont {Zhang}}, \ and\ \bibinfo {author}
  {\bibfnamefont {Y.}~\bibnamefont {Li}},\ }\bibfield  {title} {\enquote
  {\bibinfo {title} {Multiple-relaxation-time lattice {Boltzmann} kinetic model
  for combustion},}\ }\href@noop {} {\bibfield  {journal} {\bibinfo  {journal}
  {Physical Review E}\ }\textbf {\bibinfo {volume} {91}},\ \bibinfo {pages}
  {043306} (\bibinfo {year} {2015})}\BibitemShut {NoStop}%
\bibitem [{\citenamefont {Xu}\ \emph {et~al.}(2016)\citenamefont {Xu},
  \citenamefont {Zhang}, \citenamefont {Ying},\ and\ \citenamefont
  {Wang}}]{XuAG_2016}%
  \BibitemOpen
  \bibfield  {author} {\bibinfo {author} {\bibfnamefont {A.}~\bibnamefont
  {Xu}}, \bibinfo {author} {\bibfnamefont {G.}~\bibnamefont {Zhang}}, \bibinfo
  {author} {\bibfnamefont {Y.}~\bibnamefont {Ying}}, \ and\ \bibinfo {author}
  {\bibfnamefont {C.}~\bibnamefont {Wang}},\ }\bibfield  {title} {\enquote
  {\bibinfo {title} {Complex fields in heterogeneous materials under shock:
  modeling, simulation and analysis},}\ }\href {\doibase {}} {\bibfield
  {journal} {\bibinfo  {journal} {SCIENCE CHINA Physics, Mechanics \&
  Astronomy}\ }\textbf {\bibinfo {volume} {59}},\ \bibinfo {pages} {650501}
  (\bibinfo {year} {2016})}\BibitemShut {NoStop}%
\bibitem [{\citenamefont {Gan}\ \emph {et~al.}(2018)\citenamefont {Gan},
  \citenamefont {Xu}, \citenamefont {Zhang}, \citenamefont {Zhang},\ and\
  \citenamefont {Succi}}]{GanYB_2018}%
  \BibitemOpen
  \bibfield  {author} {\bibinfo {author} {\bibfnamefont {Y.}~\bibnamefont
  {Gan}}, \bibinfo {author} {\bibfnamefont {A.}~\bibnamefont {Xu}}, \bibinfo
  {author} {\bibfnamefont {G.}~\bibnamefont {Zhang}}, \bibinfo {author}
  {\bibfnamefont {Y.}~\bibnamefont {Zhang}}, \ and\ \bibinfo {author}
  {\bibfnamefont {S.}~\bibnamefont {Succi}},\ }\bibfield  {title} {\enquote
  {\bibinfo {title} {Discrete {Boltzmann} trans-scale modeling of high-speed
  compressible flows},}\ }\href {\doibase {}} {\bibfield  {journal} {\bibinfo
  {journal} {Physical Review E}\ }\textbf {\bibinfo {volume} {97}},\ \bibinfo
  {pages} {053312} (\bibinfo {year} {2018})}\BibitemShut {NoStop}%
\bibitem [{\citenamefont {Lin}\ \emph {et~al.}(2016)\citenamefont {Lin},
  \citenamefont {Xu}, \citenamefont {Zhang},\ and\ \citenamefont
  {Li}}]{LinCD_2016}%
  \BibitemOpen
  \bibfield  {author} {\bibinfo {author} {\bibfnamefont {C.}~\bibnamefont
  {Lin}}, \bibinfo {author} {\bibfnamefont {A.}~\bibnamefont {Xu}}, \bibinfo
  {author} {\bibfnamefont {G.}~\bibnamefont {Zhang}}, \ and\ \bibinfo {author}
  {\bibfnamefont {Y.}~\bibnamefont {Li}},\ }\bibfield  {title} {\enquote
  {\bibinfo {title} {Double-distribution-function discrete {Boltzmann} model
  for combustion},}\ }\href {\doibase {}} {\bibfield  {journal} {\bibinfo
  {journal} {Combustion and Flame}\ }\textbf {\bibinfo {volume} {164}},\
  \bibinfo {pages} {137--151} (\bibinfo {year} {2016})}\BibitemShut {NoStop}%
\bibitem [{\citenamefont {Scagliarini}\ \emph {et~al.}(2010)\citenamefont
  {Scagliarini}, \citenamefont {Biferale}, \citenamefont {Sbragaglia},
  \citenamefont {Sugiyama},\ and\ \citenamefont {Toschi}}]{Scagliarini_2010}%
  \BibitemOpen
  \bibfield  {author} {\bibinfo {author} {\bibfnamefont {A.}~\bibnamefont
  {Scagliarini}}, \bibinfo {author} {\bibfnamefont {L.}~\bibnamefont
  {Biferale}}, \bibinfo {author} {\bibfnamefont {M.}~\bibnamefont
  {Sbragaglia}}, \bibinfo {author} {\bibfnamefont {K.}~\bibnamefont
  {Sugiyama}}, \ and\ \bibinfo {author} {\bibfnamefont {F.}~\bibnamefont
  {Toschi}},\ }\bibfield  {title} {\enquote {\bibinfo {title} {Lattice
  {Boltzmann} methods for thermal flows: {Continuum} limit and applications to
  compressible {Rayleigh--Taylor} systems},}\ }\href {\doibase {}} {\bibfield
  {journal} {\bibinfo  {journal} {Physics of Fluids}\ }\textbf {\bibinfo
  {volume} {22}},\ \bibinfo {pages} {055101} (\bibinfo {year}
  {2010})}\BibitemShut {NoStop}%
\bibitem [{\citenamefont {Peskin}(2002)}]{peskin_2002}%
  \BibitemOpen
  \bibfield  {author} {\bibinfo {author} {\bibfnamefont {C.~S.}\ \bibnamefont
  {Peskin}},\ }\bibfield  {title} {\enquote {\bibinfo {title} {The immersed
  boundary method},}\ }\href {\doibase 10.1017/S0962492902000077} {\bibfield
  {journal} {\bibinfo  {journal} {Acta Numerica}\ }\textbf {\bibinfo {volume}
  {11}},\ \bibinfo {pages} {479--517} (\bibinfo {year} {2002})}\BibitemShut
  {NoStop}%
\bibitem [{\citenamefont {Peskin}(1977)}]{peskin_1977}%
  \BibitemOpen
  \bibfield  {author} {\bibinfo {author} {\bibfnamefont {C.~S.}\ \bibnamefont
  {Peskin}},\ }\bibfield  {title} {\enquote {\bibinfo {title} {Numerical
  analysis of blood flow in the heart},}\ }\href@noop {} {\bibfield  {journal}
  {\bibinfo  {journal} {Journal of Computational Physics}\ }\textbf {\bibinfo
  {volume} {25}},\ \bibinfo {pages} {220--252} (\bibinfo {year}
  {1977})}\BibitemShut {NoStop}%
\bibitem [{\citenamefont {Ramsden}\ and\ \citenamefont
  {Holloway}(1991)}]{Ramsden_1991}%
  \BibitemOpen
  \bibfield  {author} {\bibinfo {author} {\bibfnamefont {D.}~\bibnamefont
  {Ramsden}}\ and\ \bibinfo {author} {\bibfnamefont {G.}~\bibnamefont
  {Holloway}},\ }\bibfield  {title} {\enquote {\bibinfo {title} {Timestepping
  {Lagrangian} particles in two dimensional {Eulerian} flow fields},}\ }\href
  {\doibase {}} {\bibfield  {journal} {\bibinfo  {journal} {Journal of
  Computational Physics}\ }\textbf {\bibinfo {volume} {95}},\ \bibinfo {pages}
  {101--116} (\bibinfo {year} {1991})}\BibitemShut {NoStop}%
\bibitem [{\citenamefont {Gerya}\ and\ \citenamefont
  {Yuen}(2015)}]{Gerya_2015}%
  \BibitemOpen
  \bibfield  {author} {\bibinfo {author} {\bibfnamefont {T.~V.}\ \bibnamefont
  {Gerya}}\ and\ \bibinfo {author} {\bibfnamefont {D.~A.}\ \bibnamefont
  {Yuen}},\ }\bibfield  {title} {\enquote {\bibinfo {title} {{Rayleigh--Taylor}
  instabilities from hydration and melting propel ‘cold plumes’ at
  subduction zones},}\ }\href {\doibase {}} {\bibfield  {journal} {\bibinfo
  {journal} {Earth and Planetary Science Letters}\ }\textbf {\bibinfo {volume}
  {212}},\ \bibinfo {pages} {47--62} (\bibinfo {year} {2015})}\BibitemShut
  {NoStop}%
\bibitem [{\citenamefont {Kilkenny}\ \emph {et~al.}(1994)\citenamefont
  {Kilkenny}, \citenamefont {Glendinning}, \citenamefont {Haan}, \citenamefont
  {Hammel}, \citenamefont {Lindl}, \citenamefont {Munro}, \citenamefont
  {Remington}, \citenamefont {Weber}, \citenamefont {Knauer},\ and\
  \citenamefont {Verdon}}]{Kilkenny_1994}%
  \BibitemOpen
  \bibfield  {author} {\bibinfo {author} {\bibfnamefont {J.~D.}\ \bibnamefont
  {Kilkenny}}, \bibinfo {author} {\bibfnamefont {S.~G.}\ \bibnamefont
  {Glendinning}}, \bibinfo {author} {\bibfnamefont {S.~W.}\ \bibnamefont
  {Haan}}, \bibinfo {author} {\bibfnamefont {B.~A.}\ \bibnamefont {Hammel}},
  \bibinfo {author} {\bibfnamefont {J.~D.}\ \bibnamefont {Lindl}}, \bibinfo
  {author} {\bibfnamefont {D.}~\bibnamefont {Munro}}, \bibinfo {author}
  {\bibfnamefont {B.~A.}\ \bibnamefont {Remington}}, \bibinfo {author}
  {\bibfnamefont {S.~V.}\ \bibnamefont {Weber}}, \bibinfo {author}
  {\bibfnamefont {J.~P.}\ \bibnamefont {Knauer}}, \ and\ \bibinfo {author}
  {\bibfnamefont {C.~P.}\ \bibnamefont {Verdon}},\ }\bibfield  {title}
  {\enquote {\bibinfo {title} {A review of the ablative stabilization of the
  {Rayleigh--Taylor} instability in regimes relevant to {ICF}},}\ }\href
  {\doibase {}} {\bibfield  {journal} {\bibinfo  {journal} {Physics of
  Plasmas}\ }\textbf {\bibinfo {volume} {1}},\ \bibinfo {pages} {1379}
  (\bibinfo {year} {1994})}\BibitemShut {NoStop}%
\bibitem [{\citenamefont {Yang}\ \emph {et~al.}(2002)\citenamefont {Yang},
  \citenamefont {D’Onofrio}, \citenamefont {Kalliadasis},\ and\ \citenamefont
  {Wit}}]{Yang_2002}%
  \BibitemOpen
  \bibfield  {author} {\bibinfo {author} {\bibfnamefont {J.}~\bibnamefont
  {Yang}}, \bibinfo {author} {\bibfnamefont {A.}~\bibnamefont {D’Onofrio}},
  \bibinfo {author} {\bibfnamefont {S.}~\bibnamefont {Kalliadasis}}, \ and\
  \bibinfo {author} {\bibfnamefont {A.~D.}\ \bibnamefont {Wit}},\ }\bibfield
  {title} {\enquote {\bibinfo {title} {{Rayleigh--Taylor} instability of
  reaction-diffusion acidity fronts},}\ }\href {\doibase {}} {\bibfield
  {journal} {\bibinfo  {journal} {Journal of Chemical Physics}\ }\textbf
  {\bibinfo {volume} {117}},\ \bibinfo {pages} {9395--9408} (\bibinfo {year}
  {2002})}\BibitemShut {NoStop}%
\bibitem [{\citenamefont {Aziz}\ and\ \citenamefont
  {Chandra}(2000)}]{Aziz_2000}%
  \BibitemOpen
  \bibfield  {author} {\bibinfo {author} {\bibfnamefont {S.~D.}\ \bibnamefont
  {Aziz}}\ and\ \bibinfo {author} {\bibfnamefont {S.}~\bibnamefont {Chandra}},\
  }\bibfield  {title} {\enquote {\bibinfo {title} {Impact, recoil and splashing
  of molten metal droplets},}\ }\href {\doibase {}} {\bibfield  {journal}
  {\bibinfo  {journal} {International Journal of Heat and Mass Transfer}\
  }\textbf {\bibinfo {volume} {43}},\ \bibinfo {pages} {2841--2857} (\bibinfo
  {year} {2000})}\BibitemShut {NoStop}%
\bibitem [{\citenamefont {Peltier}\ and\ \citenamefont
  {Caulfield}(2003)}]{Peltier_2003}%
  \BibitemOpen
  \bibfield  {author} {\bibinfo {author} {\bibfnamefont {W.}~\bibnamefont
  {Peltier}}\ and\ \bibinfo {author} {\bibfnamefont {C.}~\bibnamefont
  {Caulfield}},\ }\bibfield  {title} {\enquote {\bibinfo {title} {Mixing
  efficiency in stratified shear flows},}\ }\href {\doibase {}} {\bibfield
  {journal} {\bibinfo  {journal} {Annual Review of Fluid Mechanics}\ }\textbf
  {\bibinfo {volume} {35}},\ \bibinfo {pages} {135--167} (\bibinfo {year}
  {2003})}\BibitemShut {NoStop}%
\bibitem [{\citenamefont {Zabusky}(1999)}]{Zabusky_1999}%
  \BibitemOpen
  \bibfield  {author} {\bibinfo {author} {\bibfnamefont {N.~J.}\ \bibnamefont
  {Zabusky}},\ }\bibfield  {title} {\enquote {\bibinfo {title} {Vortex paradigm
  for accelerated inhomogeneous flows: Visiometrics for the {Rayleigh--Taylor}
  and {Richtmyer--Meshkov} environments},}\ }\href {\doibase {}} {\bibfield
  {journal} {\bibinfo  {journal} {Annual Review of Fluid Mechanics}\ }\textbf
  {\bibinfo {volume} {31}},\ \bibinfo {pages} {495--536} (\bibinfo {year}
  {1999})}\BibitemShut {NoStop}%
\bibitem [{\citenamefont {Betti}\ \emph {et~al.}(1998)\citenamefont {Betti},
  \citenamefont {Goncharov}, \citenamefont {Mccrory},\ and\ \citenamefont
  {Verdon}}]{Betti_1998}%
  \BibitemOpen
  \bibfield  {author} {\bibinfo {author} {\bibfnamefont {R.}~\bibnamefont
  {Betti}}, \bibinfo {author} {\bibfnamefont {V.~N.}\ \bibnamefont
  {Goncharov}}, \bibinfo {author} {\bibfnamefont {R.~L.}\ \bibnamefont
  {Mccrory}}, \ and\ \bibinfo {author} {\bibfnamefont {C.~P.}\ \bibnamefont
  {Verdon}},\ }\bibfield  {title} {\enquote {\bibinfo {title} {Growth rates of
  the ablative {Rayleigh--Taylor} instability in inertial confinement
  fusion},}\ }\href {\doibase {}} {\bibfield  {journal} {\bibinfo  {journal}
  {Physics of Plasmas}\ }\textbf {\bibinfo {volume} {5}},\ \bibinfo {pages}
  {1446--1454} (\bibinfo {year} {1998})}\BibitemShut {NoStop}%
\bibitem [{\citenamefont {Yu}\ \emph {et~al.}(2018)\citenamefont {Yu},
  \citenamefont {Xue}, \citenamefont {Liu}, \citenamefont {Hu}, \citenamefont
  {Liu}, \citenamefont {Ye}, \citenamefont {Wang}, \citenamefont {Wu},\ and\
  \citenamefont {Fan}}]{YuCX_2018}%
  \BibitemOpen
  \bibfield  {author} {\bibinfo {author} {\bibfnamefont {C.~X.}\ \bibnamefont
  {Yu}}, \bibinfo {author} {\bibfnamefont {C.}~\bibnamefont {Xue}}, \bibinfo
  {author} {\bibfnamefont {J.}~\bibnamefont {Liu}}, \bibinfo {author}
  {\bibfnamefont {X.~Y.}\ \bibnamefont {Hu}}, \bibinfo {author} {\bibfnamefont
  {Y.~Y.}\ \bibnamefont {Liu}}, \bibinfo {author} {\bibfnamefont {W.~H.}\
  \bibnamefont {Ye}}, \bibinfo {author} {\bibfnamefont {L.~F.}\ \bibnamefont
  {Wang}}, \bibinfo {author} {\bibfnamefont {J.~F.}\ \bibnamefont {Wu}}, \ and\
  \bibinfo {author} {\bibfnamefont {Z.~F.}\ \bibnamefont {Fan}},\ }\bibfield
  {title} {\enquote {\bibinfo {title} {Multiple eigenmodes of the
  {Rayleigh--Taylor} instability observed for a fluid interface with smoothly
  varying density},}\ }\href {\doibase {}} {\bibfield  {journal} {\bibinfo
  {journal} {Physical Review E}\ }\textbf {\bibinfo {volume} {97}},\ \bibinfo
  {pages} {013102} (\bibinfo {year} {2018})}\BibitemShut {NoStop}%
\bibitem [{\citenamefont {Bernstein}\ and\ \citenamefont
  {Book}(1983)}]{Bernstein_1983}%
  \BibitemOpen
  \bibfield  {author} {\bibinfo {author} {\bibfnamefont {I.~B.}\ \bibnamefont
  {Bernstein}}\ and\ \bibinfo {author} {\bibfnamefont {D.~L.}\ \bibnamefont
  {Book}},\ }\bibfield  {title} {\enquote {\bibinfo {title} {Effect of
  compressibility on the {Rayleigh--Taylor} instability},}\ }\href {\doibase
  {}} {\bibfield  {journal} {\bibinfo  {journal} {Physics of Fluids}\ }\textbf
  {\bibinfo {volume} {26}},\ \bibinfo {pages} {453--458} (\bibinfo {year}
  {1983})}\BibitemShut {NoStop}%
\bibitem [{\citenamefont {Gauthier}\ and\ \citenamefont
  {Le~Creurer}(2010)}]{Gauthier_2010}%
  \BibitemOpen
  \bibfield  {author} {\bibinfo {author} {\bibfnamefont {S.}~\bibnamefont
  {Gauthier}}\ and\ \bibinfo {author} {\bibfnamefont {B.}~\bibnamefont
  {Le~Creurer}},\ }\bibfield  {title} {\enquote {\bibinfo {title}
  {Compressibility effects in {Rayleigh--Taylor} instability-induced flows},}\
  }\href {\doibase {}} {\bibfield  {journal} {\bibinfo  {journal}
  {Philosophical Transactions of The Royal Society A: Mathematical, Physical
  and Engineering Sciences}\ }\textbf {\bibinfo {volume} {368}},\ \bibinfo
  {pages} {1681--1704} (\bibinfo {year} {2010})}\BibitemShut {NoStop}%
\bibitem [{\citenamefont {Jin}\ \emph {et~al.}(2005)\citenamefont {Jin},
  \citenamefont {Liu}, \citenamefont {Lu}, \citenamefont {Cheng}, \citenamefont
  {Glimm},\ and\ \citenamefont {Sharp}}]{jin_2005RTI}%
  \BibitemOpen
  \bibfield  {author} {\bibinfo {author} {\bibfnamefont {H.}~\bibnamefont
  {Jin}}, \bibinfo {author} {\bibfnamefont {X.}~\bibnamefont {Liu}}, \bibinfo
  {author} {\bibfnamefont {T.}~\bibnamefont {Lu}}, \bibinfo {author}
  {\bibfnamefont {B.}~\bibnamefont {Cheng}}, \bibinfo {author} {\bibfnamefont
  {J.}~\bibnamefont {Glimm}}, \ and\ \bibinfo {author} {\bibfnamefont
  {D.}~\bibnamefont {Sharp}},\ }\bibfield  {title} {\enquote {\bibinfo {title}
  {{Rayleigh--Taylor} mixing rates for compressible flow},}\ }\href@noop {}
  {\bibfield  {journal} {\bibinfo  {journal} {Physics of Fluids}\ }\textbf
  {\bibinfo {volume} {17}},\ \bibinfo {pages} {024104} (\bibinfo {year}
  {2005})}\BibitemShut {NoStop}%
\bibitem [{\citenamefont {Xue}\ and\ \citenamefont {Ye}(2010)}]{Xue_pop_2010}%
  \BibitemOpen
  \bibfield  {author} {\bibinfo {author} {\bibfnamefont {C.}~\bibnamefont
  {Xue}}\ and\ \bibinfo {author} {\bibfnamefont {W.}~\bibnamefont {Ye}},\
  }\bibfield  {title} {\enquote {\bibinfo {title} {Destabilizing effect of
  compressibility on {Rayleigh--Taylor} instability for fluids with fixed
  density profile},}\ }\href {\doibase {}} {\bibfield  {journal} {\bibinfo
  {journal} {Physics of Plasmas}\ }\textbf {\bibinfo {volume} {17}},\ \bibinfo
  {pages} {042705} (\bibinfo {year} {2010})}\BibitemShut {NoStop}%
\bibitem [{\citenamefont {Weber}\ \emph {et~al.}(2014)\citenamefont {Weber},
  \citenamefont {Clark}, \citenamefont {Cook}, \citenamefont {Busby},\ and\
  \citenamefont {Robey}}]{Weber_pre_2014}%
  \BibitemOpen
  \bibfield  {author} {\bibinfo {author} {\bibfnamefont {C.}~\bibnamefont
  {Weber}}, \bibinfo {author} {\bibfnamefont {D.}~\bibnamefont {Clark}},
  \bibinfo {author} {\bibfnamefont {A.}~\bibnamefont {Cook}}, \bibinfo {author}
  {\bibfnamefont {L.}~\bibnamefont {Busby}}, \ and\ \bibinfo {author}
  {\bibfnamefont {H.}~\bibnamefont {Robey}},\ }\bibfield  {title} {\enquote
  {\bibinfo {title} {Inhibition of turbulence in inertial-confinement-fusion
  hot spots by viscous dissipation},}\ }\href {\doibase {}} {\bibfield
  {journal} {\bibinfo  {journal} {Physical Review E}\ }\textbf {\bibinfo
  {volume} {89}},\ \bibinfo {pages} {053106} (\bibinfo {year}
  {2014})}\BibitemShut {NoStop}%
\bibitem [{\citenamefont {Sohn}(2009)}]{Sohn_2009}%
  \BibitemOpen
  \bibfield  {author} {\bibinfo {author} {\bibfnamefont {S.-I.}\ \bibnamefont
  {Sohn}},\ }\bibfield  {title} {\enquote {\bibinfo {title} {Effects of surface
  tension and viscosity on the growth rates of {Rayleigh--Taylor} and
  {Richtmyer--Meshkov} instabilities},}\ }\href {\doibase {}} {\bibfield
  {journal} {\bibinfo  {journal} {Physical Review E}\ }\textbf {\bibinfo
  {volume} {80}},\ \bibinfo {pages} {055302} (\bibinfo {year}
  {2009})}\BibitemShut {NoStop}%
\end{thebibliography}%
\end{document}